\documentclass[a4paper,12pt]{article}
\usepackage{amssymb}
\usepackage{amsmath}
\usepackage{epsfig}

\usepackage{graphics}

\usepackage{latexsym}
\usepackage{rotating}

\title{Traveling Wave Solutions of Degenerate Coupled KdV Equation}

\author{Metin G\"{u}rses \thanks{gurses@fen.bilkent.edu.tr}\\
{\small Department of Mathematics, Faculty of Science}\\
{\small Bilkent University, 06800 Ankara - Turkey}\\
Asl{\i} Pekcan \thanks{Email:aslipekcan@hacettepe.edu.tr} \\
{\small Department of Mathematics, Faculty of Science} \\
{\small Hacettepe University, 06800 Ankara - Turkey}
}

\date{\nonumber}
\begin{document}
\maketitle
\date{\nonumber}
\newtheorem{thm}{Theorem}[section]
\newtheorem{Le}{Lemma}[section]
\newtheorem{defi}{Definition}[section]
\newtheorem{ex}{Example}[section]
\newtheorem{pro}{Proposition}[section]
\baselineskip 17pt

\numberwithin{equation}{section}

\begin{abstract}
 We give a detailed study of the traveling wave solutions of $(\ell=2)$
Kaup-Boussinesq type of coupled KdV equations. Depending upon
the zeros of a  fourth degree polynomial, we have cases where there exist
no nontrivial real solutions, cases where asymptotically decaying to a constant solitary
wave solutions, and cases where there are periodic solutions. All such possible
solutions are given explicitly in the form of Jacobi elliptic functions.
Graphs of some exact solutions in solitary wave and
periodic shapes are exhibited. Extension of our study to the cases
$\ell=3$ and $\ell=4$ are also mentioned.

\end{abstract}

\maketitle

\section{Introduction}

Multi-component Kaup-Boussinesq (KB) equations can be obtained from the Lax operator
\begin{equation}
L=D^2-\sum_{k=1}^{\ell}\lambda^{k-1}q^k(x,t),
\end{equation}
where $q^k(x,t)$, $k=1,2,...,\ell$ are the multi-KB fields \cite{alonso}-\cite{anton3}.
Here $\ell \geq 2$ is a positive integer.

The multi system of KB equation is given as
\begin{eqnarray}\label{eqn1}
\displaystyle u_t&=&\frac{3}{2}uu_x+q_x^2\nonumber \\
q_t^2&=&q^2u_x+\frac{1}{2}uq_x^2+q_x^3\nonumber \\
\vdots && \quad \vdots \quad \quad    \vdots \quad \quad \vdots\nonumber \\
q_t^{\ell-1}&=&q^{\ell-1}u_x+\frac{1}{2}uq_x^{\ell-1}+v_x \nonumber \\
v_t&=&-\frac{1}{4}u_{xxx}+v u_x+\frac{1}{2}uv_x,
\end{eqnarray}
where $q^1=u$ and $q^{\ell}=v$. This system in (\ref{eqn1}) was shown
to be also a degenerate KdV system of rank one \cite{gur1}-\cite{gurnew}. This
system admits also recursion operator for all values of $\ell$. In
this work we shall investigate the traveling wave solutions of
these coupled equations. For this purpose we start with the case
$\ell=2$. To find such solutions we use time and space
translation symmetries of the coupled system.

 The KB equation for $\ell=2$ is
\begin{eqnarray}\label{l=2}
u_t&=&\frac{3}{2}uu_x+v_x\nonumber \\
v_t&=&-\frac{1}{4}u_{xxx}+vu_x+\frac{1}{2}uv_x.
\end{eqnarray}

 In \cite{ivan} the inverse problem of the above system was studied and soliton solutions which decay asymptotically were found. The $N=1$ solution found in that work corresponds to the interaction of two solitary waves. It was also mentioned in \cite{ivan} that there is no solution in the form of traveling wave. Here in this work we prove that there exists no asymptotically vanishing traveling wave solutions of system of equations for $\ell=2$.
This is consistent with the observation of \cite{ivan}. We show that this is also valid for $\ell=4$. We claim it to be true for all even positive integers.
We show that it is possible to find solitary wave solutions of (\ref{l=2}) which asymptotically decay to non-zero constants. Furthermore in addition
to the solitary wave solutions of (\ref{l=2}) we find all traveling wave solutions which are expressible in terms of Jacobi elliptic functions.

 Traveling wave solutions of a system of equations can be obtained if the equations possess time and space translation symmetries. Such symmetries
exist in our case. Hence letting $x-ct=\xi$ where $c$ is a constant (the speed of the wave) and $u(x,t)=f(\xi)$, and $v(x,t)=g(\xi)$ from the first equation of (\ref{l=2}) we have
$$
\displaystyle -cf'=\frac{3}{2}ff'+g',
$$
which gives
\begin{equation}\label{functiong}
\displaystyle g(\xi)=-cf-\frac{3}{4}f^2+d_1,
\end{equation}
where $d_1$ is an integration constant. Using $g(\xi)$ in the second equation of (\ref{l=2}) yields
$$
\displaystyle -\frac{1}{4}f'''-3cff'-\frac{3}{2}f^2f'+(d_1-c^2)f'=0.$$
Integrating above equation once we obtain
$$
\displaystyle -\frac{1}{4}f''-\frac{3}{2}cf^2-\frac{1}{2}f^3+(d_1-c^2)f+d_2=0.$$
By using $f'$ as an integrating factor, we can integrate once more. Finally we get
\begin{equation}\label{F(f)}
(f')^2=-f^4-4cf^3+4(d_1-c^2)f^2+8d_2f+8d_3=F(f),
\end{equation}
where $c, d_1, d_2, d_3$ are constants. These constants can be determined from the initial conditions $f(0)$, $f'(0)$, $f''(0)$ and $g(0)$. If $F(f)$ has zeros, these zeros are related to these initial conditions. For asymptotically decaying solutions of $(\ell=2)$ KB equations $f$, $f'$, $f''$, $f'''$, $g$, and $g'$ go to zero as $\xi\rightarrow \pm \infty$.  Here in this work we shall find all possible solutions $f$ of (\ref{F(f)}). Given a solution $f$ one can find the corresponding solution $g(\xi)$ from (\ref{functiong}).

 In \cite{pav} and \cite{el}, a KB  like system
\begin{eqnarray}
&&h_{t}+(uh)_{x}+\frac{1}{4}\,u_{xxx}=0,\nonumber\\
&&u_{t}+uu_{x}+h_{x}=0,\label{kb2}
\end{eqnarray}
was considered. Traveling wave solutions of this system satisfy a differential equation like (\ref{F(f)}) but the corresponding polynomial
$F_1(f)$ is asymptotically positive definite. This means that the above KB like system possesses asymptotically decaying traveling wave solutions. In \cite{pav} and \cite{el}  some solitary wave solutions were found. The fourth degree polynomial arising in traveling wave solutions of the system (\ref{kb2}) is different than the one given in (\ref{F(f)}). Hence the behavior of solutions here in this work  and in Refs. \cite{pav} and \cite{el} are different.

 In \cite{kamc}  a modified version of the system (\ref{kb2}), i.e.
\begin{eqnarray}
&&h_{t}+(uh)_{x} \pm \frac{1}{4}\, \varepsilon^2\, u_{xxx}=0,\nonumber\\
&&u_{t}+uu_{x}+h_{x}=0,\label{kb4}
\end{eqnarray}
was considered, where $\varepsilon$ is a parameter which controls the dispersion effects. The upper sign is for the case when the gravity force dominates over the capillary one, and the lower sign is for the opposite case when capillary dominates over the gravity. The traveling wave solutions of the above system (\ref{kb4}) were considered in \cite{kamc}. The equation (\ref{F(f)}) becomes now
$ \varepsilon^2\,(f^{\prime})^2=\pm F_2(f)$. In both cases solitary wave solutions (dark and bright solitons) were found in \cite{kamc}.
The lower case (negative sign) resembles to our case. Hence our solution in section $3.1$ can be considered as a dark soliton in the sense of \cite{kamc}.
This is the solution corresponding one double and two simple zeros of the polynomial $F(f)$. We have all other solutions corresponding to different combinations of the zeros of $F(f)$ in sections $3$, $4$, and $5$.

The layout of our paper is as follows: In section $2$, we study the behavior of the solutions in the neighborhood of the zeros of $F(f)$ and discuss all possible cases. We find all
solitary wave solutions of the system (\ref{l=2}) in section $3$. These correspond to one double and two simple zeros of $F(f)$, and one triple and one simple zeros of $F(f)$. In section $4$, we find all elliptic type of solutions starting from very special ones to the most general elliptic type of solutions. These solutions are given in terms of the zeros of the function $F(f)$.
In section $5$, we discuss $\ell=3$ and $\ell=4$ cases. In section $6$, we give the graphs of the solutions corresponding to all cases considered in the text.

\section{General waves of permanent form for $(\ell=2)$}

\begin{pro} There is no real asymptotically vanishing traveling wave solution of the equation (\ref{l=2}) in the
form $u(x,t)=f(\xi)$ and $v(x,t)=g(\xi)$, where $\xi=x-ct$.
\end{pro}

\noindent \textbf{Proof.} If we apply the boundary conditions
$f,f',f'',f''', g, g' \rightarrow 0$ as $\xi\rightarrow \pm \infty$ which
describe the solitary wave, we get $d_1=d_2=d_3=0$. Hence we end
up with
\begin{eqnarray*}
(f')^2=-f^4-4cf^3-4c^2f^2&=&-f^2(f^2+4cf+4c^2)\\
&=&-f^2(f+2c)^2.
\end{eqnarray*}
Clearly, we do not have a real solution $f$.
\hfill $\Box$

Now we will deal with the equation (\ref{F(f)}). In order to have real solutions,
$d_1,d_2,d_3$ must take values so that the following inequality holds:
$$
4d_1f^2+8d_2f+8d_3\geq f^2(f+2c)^2.$$

\bigskip

\subsection{Zeros of $F(f)$ and Types of Solutions}

Here we will analyze the zeros of $F(f)$.

\noindent \textbf{(i)} \, If $f_1=f(\xi_1)$ is a \textit{simple zero} of $F(f)$ we have $F(f_1)=0$. Taylor
expansion of $F(f)$ gives
\begin{eqnarray*}
(f')^2=F(f)&=&F(f_1)+F'(f_1)(f-f_1)+O((f-f_1)^2)\\
&=&F'(f_1)(f-f_1)+O((f-f_1)^2).
\end{eqnarray*}
From here we get $f'(\xi_1)=0$ and $\displaystyle f''(\xi_1)=F'(f_1)/2$. Hence we can write the function $f(\xi)$ as
\begin{eqnarray}
\displaystyle f(\xi)&=&f(\xi_1)+(\xi-\xi_1)f'(\xi_1)+\frac{1}{2}(\xi-\xi_1)^2f''(\xi_1)+O((\xi-\xi_1)^3)\nonumber\\
&=& f_1+\frac{1}{4}(\xi-\xi_1)^2F'(f_1)+O((\xi-\xi_1)^3).
\end{eqnarray}
Thus, in the neighborhood of $\xi=\xi_1$, the function $f(\xi)$ has local minimum or maximum as $F'(f_1)$ is positive or negative respectively since
$\displaystyle f''(\xi_1)=F'(f_1)/2$.\\
\bigskip

\noindent \textbf{(ii)} \, If $f_1=f(\xi_1)$ is a \textit{double zero} of $F(f)$ we have $F(f_1)=F'(f_1)=0$. Taylor
expansion of $F(f)$ gives
\begin{eqnarray}\label{doublezero}
\displaystyle (f')^2=F(f)&=&F(f_1)+F'(f_1)(f-f_1)+\frac{1}{2}(f-f_1)^2F''(f_1)+O((f-f_1)^3)\nonumber \\
&=&\frac{1}{2}(f-f_1)^2F''(f_1)+O((f-f_1)^3).
\end{eqnarray}
To have real solution $f$, we should have $F''(f_1)> 0$. From the equality (\ref{doublezero}) we get
$$
\displaystyle f'\pm \frac{1}{\sqrt{2}}f\sqrt{F''(f_1)}\sim \pm \frac{1}{\sqrt{2}}f_1\sqrt{F''(f_1)},$$
which gives
\begin{equation}
\displaystyle f(\xi) \sim  f_1+\alpha e^{\pm \frac{1}{\sqrt{2}}\sqrt{F''(f_1)}\xi},
\end{equation}
where $\alpha$ is a constant. Hence $f\rightarrow f_1$ as $\xi\rightarrow \mp \infty$. The solution $f$ can have only one peak and the wave extends
from $-\infty$ to $\infty$.\\
\bigskip

\noindent \textbf{(iii)} \, If $f_1=f(\xi_1)$ is a \textit{triple zero} of $F(f)$ we have $F(f_1)=F'(f_1)=F''(f_1)=0$. Taylor
expansion of $F(f)$ gives
\begin{eqnarray}\label{triplezero}
\displaystyle (f')^2&=&F(f)\nonumber\\
&=&F(f_1)+F'(f_1)(f-f_1)+\frac{1}{2}(f-f_1)^2F''(f_1)+\frac{1}{6}(f-f_1)^3+O((f-f_1)^4)\nonumber \\
&=&\frac{1}{6}(f-f_1)^3F'''(f_1)+O((f-f_1)^4).
\end{eqnarray}
This is valid only if both signs of $(f-f_1)^3$ and $F'''(f_1)$ are same i.e. we have the following two possibilities to have real solution $f$:

\noindent $1)$ $(f-f_1)> 0$ and $F'''(f_1)>0,$ \\

\noindent $2)$ $(f-f_1)< 0$ and $F'''(f_1)<0.$\\

\noindent Let us analyze these cases. If $(f-f_1)> 0$ and $F'''(f_1)>0$ then we have
$$
\displaystyle f'\sim \pm \frac{1}{\sqrt{6}}(f-f_1)^{3/2}\sqrt{F'''(f_1)},$$
which gives
\begin{equation}
\displaystyle f(\xi)\sim f_1+\frac{4}{\Big(\pm \frac{1}{\sqrt{6}}\sqrt{F'''(f_1)}\xi+\alpha_1\Big)^2},
\end{equation}
where $\alpha_1$ is a constant. Thus $f\rightarrow f_1$ as $\xi\rightarrow \pm \infty$ if $F'''(f_1)>0$.

\noindent Let $(f-f_1)< 0$ and $F'''(f_1)<0$ hold. In this case, $(f_1-f)> 0$ and $F'''(f_1)=-G(f_1)$, $G(f_1)>0$. Then
$$
\displaystyle f'\sim \pm \frac{1}{\sqrt{6}}(f_1-f)^{3/2}\sqrt{G(f_1)},$$
which yields
\begin{equation}
\displaystyle f(\xi)\sim f_1-\frac{4}{\Big(\pm \frac{1}{\sqrt{6}}\sqrt{G(f_1)}\xi+\alpha_2\Big)^2},
\end{equation}
where $\alpha_2$ is a constant. Thus $f\rightarrow f_1$ as $\xi\rightarrow \pm \infty$ if $F'''(f_1)=-G(f_1)<0$.

\bigskip

\noindent \textbf{(iv)} \,If $f_1=f(\xi_1)$ is a \textit{quadruple zero} of $F(f)$ then there is only one possibility $F(f)=-(f-f_1)^4=(f')^2$. It is clear that this case does not give a real solution except when $f=f_1$.

\subsection{All Possible Cases}

\noindent Here we present the sketches of the graphs of $F(f)$. Real solutions $(f')^2=F(f)\geq 0$ occur in the shaded regions.
\begin{figure}[!h]
\centering
\includegraphics[angle=0,scale=.3]{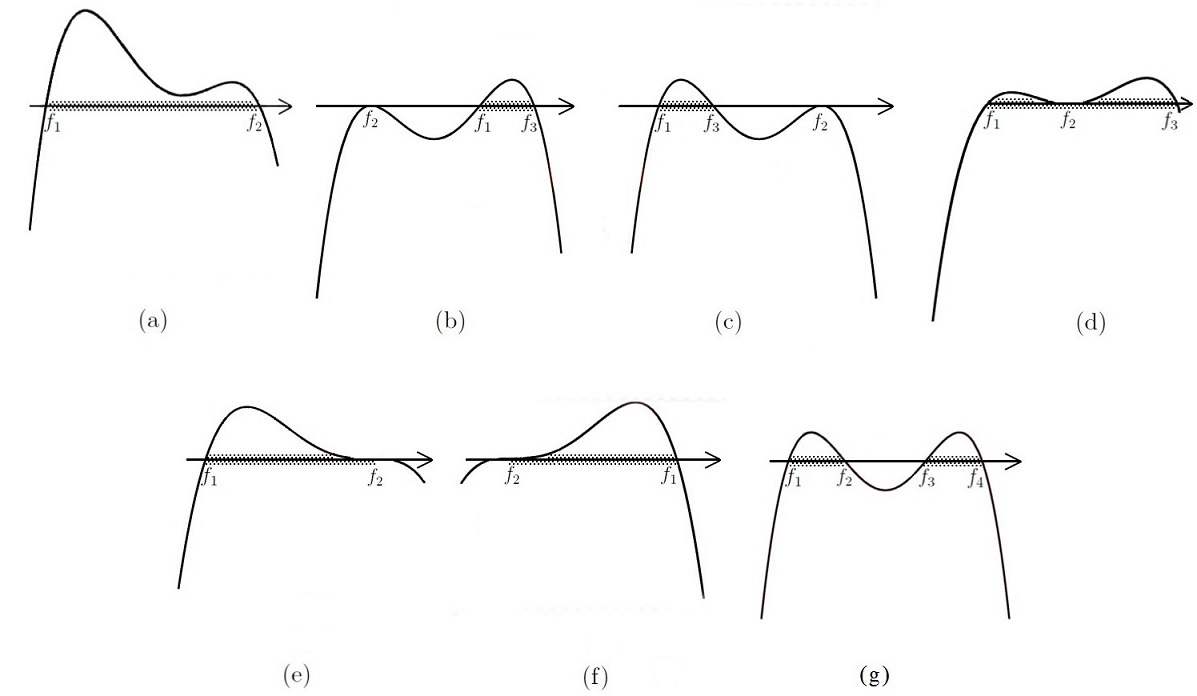}
\caption{All possible sketches of the graphs of $F(f)$}
\end{figure}

Now we analyze all possible cases about the zeros of $F(f)$ and above graphs.

\noindent\textbf{(1)\, No real zero.} If there is no real zeros of $F(f)$ then $F(f)<0$.
Hence there is no real solution of (\ref{F(f)}) in that case.

\noindent \textbf{(2)\, Two simple real zeros.} If there is a simple zero $f_1$ of $F(f)$, since the order of $F(f)$ is four,
there should be another simple zero $f_2$ of $F(f)$. The corresponding graph to this case is given in $(a)$.
Here, the real solution occurs when $f$ is between two different simple zeros $f_1$ and $f_2$. At $f_1$, $F'(f_1)=f''(\xi_1)> 0$
so graph of the function $f$ is concave up at $\xi_1$. At $f_2$, $F'(f_2)=f''(\xi_2)> 0$
hence graph of the function $f$ is concave up at $\xi_2$. Thus it is clear that the solution is periodic.

\noindent \textbf{(3)\, One double zero.} If there is only one double zero $f_1$
then
\begin{equation}
(f')^2=-(f-f_1)^2(f^2+pf+q),
\end{equation}
where $f^2+pf+q$ has no real zero. This means $p^2-4q< 0$ which yields that $f^2+pf+q>0$. Then $(f')^2=-(f-f_1)^2(f^2+pf+q)<0$ hence there is no real solution in that case except when $f=f_1$.
Similarly, in the case when $F(f)$ has two double zeros $f_1$ and $f_2$, no real solutions exist since
$(f')^2=-(f-f_1)^2(f-f_2)^2<0$ except when $f=f_1$ or $f=f_2$.

\noindent \textbf{(4)\, One double and two simple zeros.} The corresponding graphs for this case are $(b)$, $(c)$ and $(d)$.
In $(b)$ and $(c)$, there are two different simple zeros $f_1$ and $f_3$ and one double zero $f_2$. We have $f_2<f_1<f_3$ in $(b)$ and in the graph $(c)$, $f_1<f_3<f_2$. In both cases, the real solution occurs when $f$ is between two simple zeros $f_1$ and $f_3$. At $f_1$, $F'(f_1)=f''(\xi_1)> 0$ so graph of the function $f$ is concave up at $\xi_1$. At $f_3$, $F'(f_3)=f''(\xi_3)> 0$ hence graph of the function $f$ is concave up at $\xi_3$. It is clear that the solution is periodic in this case.

In $(d)$, different than the graphs $(b)$ and $(c)$ we have $f_1<f_2<f_3$. The real solution occurs when $f$ stays between
$f_1$ and $f_2$ or $f_2$ and $f_3$. At $f_1$, $F'(f_1)=f''(\xi_1)> 0$ hence graph of the function $f$ is concave up at $\xi_1$. At double zero $f_2$,
$f\rightarrow f_2$ as $\xi \rightarrow \pm \infty$. Hence we have a solitary wave solution with amplitude $f_1-f_2<0$.

Similarly at $f_3$, $F'(f_3)=f''(\xi_3)<0$, hence graph of the function $f$ is concave down at $\xi_3$. Therefore, we also have a solitary wave solution with amplitude $f_3-f_2>0$. Explicit solitary wave solution for this case can be found in the next section.

\noindent \textbf{(5) \, One triple and one simple zero.} For this case, we can analyze the graphs $(e)$ and $(f)$. In $(e)$, $f_1$ is simple and $f_2$ is triple zeros of $F(f)$. We see that $F'(f_1)=f''(\xi_1)>0$ hence graph of the function $f$ is concave up at $\xi_1$. From the case (iii), we know that $f\rightarrow f_2$ as $\xi \rightarrow \pm \infty$ for $f-f_2 < 0$ and $F'''(f_2)< 0$. Hence we have solitary wave solution with amplitude $f_1-f_2< 0$.

Similarly, in $(f)$ we have one triple zero $f_1$ and one simple zero $f_2$. For triple zero $f_1$ we have $f\rightarrow f_1$ as $\xi \rightarrow \pm \infty$ for $f-f_1 > 0$ and $F'''(f_2)> 0$. For simple zero we have $F'(f_2)=f''(\xi_2)< 0$ therefore graph of the function $f$ is concave down at $\xi_2$. Clearly, we have a solitary wave solution with amplitude $f_2-f_1> 0$. Explicit solitary wave solution for this case can be found in the next section.

\noindent \textbf{(6)\, Four different simple zeros.} The corresponding graph for this case is given in $(g)$. Here, there are four simple zeros $f_1<f_2<f_3<f_4$. For $f_1$ and $f_3$, we have $F'(f_1)=f''(\xi_1)>0$ and $F'(f_3)=f''(\xi_3)>0$ thus graph of the function $f$ is concave up at $\xi_1$ and $\xi_3$. For $f_2$ and $f_4$, we have $F'(f_2)=f''(\xi_2)<0$ and $F'(f_4)=f''(\xi_4)<0$ so graph of
the function $f$ is concave down at $\xi_2$ and $\xi_4$. Obviously, the solution is periodic.

\noindent As a summary we have the following results. By solution below, we mean non-constant solutions.
 \begin{pro}\label{pro22} Equation (\ref{F(f)}) has no real solutions when the function F(f) has one of the following properties: (i) it has no real zeros, (ii) it has only two real zeros, (iii) it has only one double zero, (iv) it has only two double zeros, and (v) it has a quadruple zero.
\end{pro}
\begin{pro}
Equation (\ref{F(f)}) admits solitary wave solutions when the function F(f) admits (i) one double  and two simple zeros and (ii) one triple and one simple zeros.
\end{pro}

\noindent From the proposition \ref{pro22} we can conclude that the function $F(f)$ must have four zeros,
$$
F(f)=-(f-f_1)(f-f_2)(f-f_3)(f-f_4).$$
The constants $c, d_1, d_2, d_3$ can be expressed in terms of the zeros of $F(f)$:
\begin{eqnarray}\label{parameters}
\displaystyle
c&=&-\frac{f_1+f_2+f_3+f_4}{4}\nonumber\\
d_1&=&\frac{(f_1+f_2+f_3+f_4)^2}{16}-\frac{f_1f_2+f_2f_4+f_2f_3+f_1f_4+f_1f_3+f_3f_4}{4}\nonumber\\
d_2&=&\frac{f_1f_2f_4+f_1f_2f_3+f_2f_3f_4+f_1f_3f_4}{8}\nonumber\\
d_3&=&-\frac{f_1f_2f_3f_4}{8}.
\end{eqnarray}

In the next section we shall find the solitary wave solutions mentioned in the above proposition which correspond to special cases of the zeros $f_1, f_2, f_3, f_4$.

\bigskip

\section{Exact Solitary Wave Solutions}

\subsection{One double zero and two simple zeros}

\noindent Let $f_1$ and $f_3$ be two different simple zeros and $f_2$ be a double zero of $F(f)$. Thus we have
$$
(f')^2=F(f)=-(f-f_2)^2(f-f_1)(f-f_3).$$
Let $f-f_2=u$ and so $f-f_1=u-u_1$, where $u_1=f_1-f_2$ and $f-f_2=u-u_3$, where $u_3=f_3-f_2$. Hence the above equation becomes
$$
(u')^2=-u^2(u-u_1)(u-u_3).
$$
Using the substitution $\displaystyle u=1/y$
\begin{eqnarray*}
\displaystyle (y')^2=-y^2\Big(\frac{1}{y}-u_1\Big)\Big(\frac{1}{y}-u_3\Big)&=&-(1-yu_1)(1-yu_3)\\
&=&-u_1u_3\Big(\frac{1}{u_1}-y\Big)\Big(\frac{1}{u_3}-y\Big).
\end{eqnarray*}
After some arrangements we have
\begin{equation}\label{doublesimplesimple}
\displaystyle (y')^2=-u_1u_3\Big\{\Big[y-\frac{1}{2}\Big(\frac{1}{u_1}+\frac{1}{u_3}\Big)\Big]^2-\frac{1}{4}\Big(\frac{1}{u_1}-\frac{1}{u_3}\Big)^2\Big\}.
\end{equation}
Using the trigonometric substitution
$$
\displaystyle y-\frac{1}{2}\Big(\frac{1}{u_1}+\frac{1}{u_3}\Big)=\frac{1}{2}\Big(\frac{1}{u_1}-\frac{1}{u_3}\Big)\cosh{\theta}
$$
the equation (\ref{doublesimplesimple}) becomes
$$
(\theta')^2=-u_1u_3.
$$
Note that in the case when $F(f)$ has two different simple zeros and one double zero, the solitary wave solution occurs only
when we have $f_1<f_2<f_3$ and this makes $u_1u_3<0$ or $-u_1u_3>0$. So from the above equation we get $\theta'=\pm\sqrt{-u_1u_3}$
which yields
$$
\theta=\pm \sqrt{-u_1u_3}(\xi-\xi_0),
$$
 where $\xi_0$ is an integration constant. Hence the solution $f$ is
 \begin{equation}\label{onedoubletwosimple}
 \displaystyle f=f_2+\frac{2}{c_1+c_2\cosh({\sqrt{(f_2-f_1)(f_3-f_2)}(\xi-\xi_0)})},
 \end{equation}
 where $ \displaystyle c_1=\Big(\frac{1}{f_1-f_2}
 +\frac{1}{f_3-f_2}\Big)$ and $ \displaystyle c_2=\Big(\frac{1}{f_1-f_2}-\frac{1}{f_3-f_2}\Big)$. It is clear that $f\rightarrow f_2$ as $\xi\rightarrow \pm \infty$.

 Note that when $u_1u_3>0$ which means $f_1<f_3<f_2$ or $f_2<f_1<f_3$ we have the following solution which is not a solitary wave solution:
 \begin{equation}
 \displaystyle f=f_2+\frac{2}{c_1\pm c_2\sin({\sqrt{(f_2-f_1)(f_3-f_2)}(\xi-\xi_0)})},
 \end{equation}
with the same $c_1$ and $c_2$ stated above.

\subsection{One triple zero and one simple zero}

\noindent Let $f_1$ be simple and $f_2$ be triple zeros of $F(f)$. Hence
\begin{equation}\label{triplesimple}
(f')^2=F(f)=-(f-f_1)^3(f-f_2).
\end{equation}
The relations between the zeros of $F(f)$ and the parameters are
\begin{equation}
\displaystyle c=-\frac{f_2+3f_1}{4},\, d_1=\frac{f_2^2-6f_1f_2-3f_1^2}{16},\, d_2=\frac{3f_1^2f_2+f_1^3}{8},\, d_3=-\frac{f_1^3f_2}{8}.
\end{equation}

\noindent Let us solve the equation (\ref{triplesimple}). Let $f-f_1=u$ so (\ref{triplesimple}) becomes
$$
(u')^2=-u^3(u-u_0), \quad u_0=f_2-f_1.
$$
We have
$$
\displaystyle \frac{du}{u^{3/2}\sqrt{u_0-u}}=\frac{\sqrt{u_0-u}\, du}{(u_0-u)u\sqrt{u}}=d\xi.
$$
By making the substitution $\displaystyle t=\sqrt{u_0-u}/\sqrt{u}$, the above equality can be solved as
$$
\displaystyle \frac{-2}{u_0}\sqrt{\frac{u_0-u}{u}}=\xi-\xi_0,
$$
where $\xi_0$ is an integration constant. Hence we find
$$
\displaystyle u=\frac{u_0}{1+\frac{u_0^2}{4}(\xi-\xi_0)^2},
$$
and inserting $u=f-f_1$ and $u_0=f_2-f_1$ we get the solution
$$
\displaystyle f=f_1+\frac{f_2-f_1}{1+\frac{1}{4}(f_2-f_1)^2(\xi-\xi_0)^2}.
$$
It is clear that $f\rightarrow f_1$ as $\xi\rightarrow\pm \infty$.

\subsection{Limiting Cases}

Here we will analyze the solution (\ref{onedoubletwosimple}) which corresponds to the case when $F(f)$ has
one double zero $f_2$ and two different simple zeros $f_1$ and $f_3$.

\noindent \textbf{(a)} When $f_1+f_3=2f_2$, the solution (\ref{onedoubletwosimple}) reduces to
 \begin{equation}\label{limitinga}
 \displaystyle f=f_2+\frac{2(f_2-f_1)(f_3-f_2)}{(f_3-f_1)}\mathrm{sech}(\sqrt{(f_2-f_1)(f_3-f_2)}(\xi-\xi_0)).
 \end{equation}

\bigskip

\noindent \textbf{(b)} When $2f_1f_3=f_2(f_1+f_3)$, the solution (\ref{onedoubletwosimple}) reduces to
\begin{equation}
 \displaystyle f=\frac{c_2\cosh(\sqrt{(f_2-f_1)(f_3-f_2)}(\xi-\xi_0))}{c_1+c_2\cosh(\sqrt{(f_2-f_1)(f_3-f_2)}(\xi-\xi_0))},
\end{equation}\label{limitingb}
which can be converted to
\begin{equation}
 \displaystyle f=\frac{c_2}{c_2+c_1\mathrm{sech}(\sqrt{(f_2-f_1)(f_3-f_2)}(\xi-\xi_0))},
\end{equation}
 where $ \displaystyle c_1=\Big(\frac{1}{f_1-f_2}
 +\frac{1}{f_3-f_2}\Big)$ and $ \displaystyle c_2=\Big(\frac{1}{f_1-f_2}-\frac{1}{f_3-f_2}\Big)$.

 \bigskip

\noindent \textbf{(c)} When $f_2=0$, then the solution (\ref{onedoubletwosimple}) reduces to
\begin{equation}
 \displaystyle f=\frac{2f_1f_3}{(f_1+f_3)+(f_3-f_1)\cosh(\sqrt{-f_1f_3}(\xi-\xi_0))},
\end{equation}
which can also be written as
\begin{equation}
 \displaystyle f=\frac{2f_1f_3\mathrm{sech}(\sqrt{-f_1f_3}(\xi-\xi_0))}{(f_3-f_1)+(f_1+f_3)\mathrm{sech}(\sqrt{-f_1f_3}(\xi-\xi_0))}, \quad f_1f_3<0.
\end{equation}

\bigskip

\noindent\textbf{(d)} When $f_2\rightarrow f_1$ or $f_2\rightarrow f_3$, then the case turns to the case when
$F(f)$ has one triple zero and one simple zero. If $f_2\rightarrow f_1$, the solution (\ref{onedoubletwosimple}) reduces to $f=f_1$,
and if $f_2\rightarrow f_3$, the solution (\ref{onedoubletwosimple}) reduces to
\begin{equation}\label{limitingd2}
\displaystyle f=f_3+\frac{f_1-f_3}{1+\frac{1}{4}(f_1-f_3)^2(\xi-\xi_0)^2}.
\end{equation}

\section{Exact Solutions in Terms of Elliptic Functions }

In this section we will find exact solutions of (\ref{l=2}) by using the Jacobi elliptic functions \cite{bradbury}. Let us give the list of the Jacobi elliptic functions and first order differential equations
satisfied by them.

\subsection{Jacobi Elliptic Functions}
\begin{eqnarray}
&&\quad y=\mathrm{sn}v \quad  (y')^2=(1-y^2)(1-k^2y^2),\label{eq:jacobisnu} \\
&&\quad y=\mathrm{cn}v \quad  (y')^2=(1-y^2)(1-k^2+k^2y^2),\label{eq:jacobicnu}\\
&&\quad y=\mathrm{dn}v \quad  (y')^2=(1-y^2)(y^2-1+k^2),\label{eq:jacobidnu}\\
&&\quad y=\mathrm{tn}v \quad  (y')^2=(1+y^2)[1+(1-k^2)y^2],\label{eq:jacobitnu}\\
&&\quad \displaystyle y=\frac{1}{\mathrm{sn}v} \quad (y')^2=(y^2-1)(y^2-k^2),\label{eq:jacobioneoversnu}\\
&&\quad \displaystyle y=\frac{1}{\mathrm{cn}v}  \quad (y')^2=(y^2-1)[(1-k^2)y^2+k^2],\label{eq:jacobioneovercnu}\\
&&\quad y=\mathrm{dn}v\mathrm{tn}v \quad (y')^2=(1+y^2)^2-4k^2y^2,\label{eq:jacobidnutnu}
\end{eqnarray}
and for the squares of these functions we have cubic equations
\begin{eqnarray}
&& \quad y=\mathrm{sn}^2v \quad (y')^2=4y(1-y)(1-k^2y),\label{eq:jacobisquaresn}\\
&& \quad y=\mathrm{cn}^2v \quad (y')^2=4y(1-y)(1-k^2+k^2y),\label{eq:jacobisquarecn}\\
&& \quad y=\mathrm{dn}^2v \quad (y')^2=4y(1-y)(y-1+k^2),\label{eq:jacobisquaredn}\\
&& \quad y=\mathrm{tn}^2v \quad (y')^2=4y(1+y)[1+(1-k^2)y],\label{eq:jacobisquaretn}\\
&& \quad y=\frac{1}{\mathrm{cn}^2v} \quad (y')^2=4y(y-1)[(1-k^2)y+k^2],\label{eq:jacobisquareoneovercn}\\
&& \quad \displaystyle y=\frac{1}{\mathrm{sn}^2v} \quad (y')^2=4y(y-1)[y-k^2],\label{eq:jacobisquareoneoversn}\\
&& \quad y=\mathrm{dn}^2v\mathrm{tn}^2v \quad (y')^2=4y[(1+y)^2-4k^2y].\label{eq:jacobisquaredntn}
\end{eqnarray}
We will also make analysis at the limiting points $k=0$ and $k=1$. Remind that
\begin{eqnarray}\label{ellipticlimits}
&&k=0 \quad \mathrm{sn}v=\sin v \quad , \quad \mathrm{cn}v =\cos v \quad , \quad \mathrm{dn} v=1,\nonumber\\
&& k=1 \quad \mathrm{sn}v=\tanh v \quad , \quad \mathrm{cn}v =\mathrm{dn}v =\mathrm{sech}v.
\end{eqnarray}

\subsection{Special Solutions of (\ref{l=2}) in Terms of Elliptic Functions}

\noindent For some special values of $c,d_1,d_2,d_3$, we have  solutions of (\ref{l=2}) in terms of Jacobi elliptic functions.
Here we will present two such types of solutions.

\noindent \textbf{Case 1.}\quad \textbf{Solutions of the form $u(x,t)=f(\xi)=\gamma+\alpha y(\beta\xi)$}

Here we shall find the solutions of (\ref{l=2}) having the form $u(x,t)=f(\xi)=\gamma+\alpha y(\beta\xi)$, where $\gamma, \alpha, \beta$ are constants, $\xi=x-ct$ and $y$ is one of the Jacobi elliptic functions. When we use this form in (\ref{F(f)}) we get the following equation:
\begin{eqnarray}\label{exactsolncaseone}
\displaystyle (y')^2&=&-\frac{\alpha^2}{\beta^2}y^4-\frac{4\alpha}{\beta^2}(c+\gamma)y^3+\frac{2}{\beta^2}(2d_1-6c\gamma-3\gamma^2-2c^2)y^2\nonumber\\
&&+\frac{4}{\alpha\beta^2}(2d_2+2d_1\gamma-2c^2\gamma-3c\gamma^2-\gamma^3)y\nonumber\\
&&+\frac{1}{\alpha^2\beta^2}(-\gamma^4-4c^2\gamma^2+8d_2\gamma-4c\gamma^3+8d_3+4d_1\gamma^2).
\end{eqnarray}
Since the parameters are real, we have $\alpha^2/\beta^2> 0$. Hence the coefficient of the term $y^4$ is negative. Thus there are two possibilities: $\alpha^2=k^2\beta^2$ which corresponds to Jacobi elliptic
function $\mathrm{cn}v$ and $\alpha^2=\beta^2$ corresponding to $\mathrm{dn}v$. Comparing the differential equations for $\mathrm{cn}v$ and $\mathrm{dn}v$ with (\ref{exactsolncaseone}), we note that the coefficients
of the terms $y^3$ and $y$ should be zero. That gives
\begin{eqnarray*}
\displaystyle \gamma&=&-c=\frac{f_1+f_2+f_3+f_4}{4}\\
d_2&=&cd_1\\
&=&\frac{f_1+f_2+f_3+f_4}{64}\Big[4(f_1f_2+f_1f_3+f_1f_4+f_2f_3+f_2f_4+f_3f_4)\\
&&-(f_1+f_2+f_3+f_4)^2\Big],
\end{eqnarray*}
where $d_1$ is given in (\ref{parameters}). Note that the equality $d_2=cd_1$ yields a relation between the zeros of
$F(f)$:
\begin{equation}\label{relationzeros}
(f_1+f_2-f_3-f_4)(f_1+f_3-f_2-f_4)(f_1+f_4-f_2-f_3)=0.
\end{equation}
The equation \eqref{exactsolncaseone} is simplified as
\begin{equation}
\displaystyle (y')^2=-\frac{\alpha^2}{\beta^2}y^4+\frac{\mu_2}{\beta^2}y^2+\frac{\mu_0}{\alpha^2\beta^2},
\end{equation}
where
\begin{eqnarray*}
\displaystyle \mu_2&=&\frac{3}{8}(f_1+f_2+f_3+f_4)^2-(f_1f_2+f_1f_3+f_1f_4+f_2f_3+f_2f_4+f_3f_4)\\
\mu_0&=&\frac{(f_1+f_2+f_3+f_4)^2}{16}(f_1f_2+f_1f_3+f_1f_4+f_2f_3+f_2f_4+f_3f_4)\\
&&-\frac{5}{256}(f_1+f_2+f_3+f_4)^4-f_1f_2f_3f_4.
\end{eqnarray*}
Here we shall take $f_4=f_1+f_3-f_2$ since the equation (\ref{relationzeros}) should be satisfied. Note that we cannot have
$f_1=f_2$ which gives $f_3=f_4$ implying two double zeros that in this case we do not have real solution $f$. This is also same in the case
of $f_2=f_3$. Hence in the below computations $f_1\neq f_2$ and $f_2\neq f_3$.

\bigskip

\noindent \textbf{1.a} \quad \textbf{cn solution}\\
Let $y=\mathrm{cn}(\beta\xi)$ with $\xi=x-ct$ where the function $y$ satisfies the first order differential equation \eqref{eq:jacobicnu}. Hence when we compare the coefficients of (\ref{exactsolncaseone}) and \eqref{eq:jacobicnu},
we get
$$
\displaystyle \beta^2=\frac{\mu_2}{2k^2-1}, \quad
\alpha^2=\frac{k^2\mu_2}{2k^2-1},\quad
k^2=\frac{1}{2}+\frac{\mu_2}{2\sqrt{4\mu_0+\mu_2^2}}.
$$
Explicitly we have
\begin{equation}\label{cnsolnparameters1}
\beta^2=(f_2-f_3)(f_1-f_2),\quad \alpha^2=(f_1-f_3)^2/4,\quad k^2=(f_1-f_3)^2/4(f_2-f_3)(f_1-f_2),
\end{equation}
or
\begin{eqnarray}\label{cnsolnparameters2}
&&\beta^2=(f_3-f_2)(f_1-f_2),\quad \alpha^2=(f_1+f_3-2f_2)^2/4,\nonumber\\
 &&k^2=(f_1+f_3-2f_2)^2/4(f_3-f_2)(f_1-f_2).
\end{eqnarray}
\noindent Hence the corresponding solution is
\begin{equation}\label{case1afirstsoln}
u(x,t)=\pm \frac{(f_1-f_3)}{2}\mathrm{cn}\Big[\sqrt{(f_2-f_3)(f_1-f_2)}\Big(x+\frac{f_1+f_3}{2}t\Big)\Big]+\frac{f_1+f_3}{2}
\end{equation}
or
\begin{equation}\label{case1asecondsoln}
u(x,t)=\pm \frac{(f_1+f_3-2f_2)}{2}\mathrm{cn}\Big[\sqrt{(f_2-f_3)(f_1-f_2)}\Big(x+\frac{f_1+f_3}{2}t\Big)\Big]+\frac{f_1+f_3}{2}.
\end{equation}
\noindent Let us check the limiting points. It is enough to consider the parameters (\ref{cnsolnparameters1}) and the solution
  (\ref{case1afirstsoln}). We can analyze (\ref{cnsolnparameters2}) and (\ref{case1asecondsoln}) similarly. For $k=0$, we have $f_1=f_3$ and the solution becomes $u(x,t)=f_1$. For $k=1$, we get the relation
\begin{equation}
2f_2=f_1+f_3.
\end{equation}
Hence the solution is
\begin{equation}\label{case1asoln}
u(x,t)=\pm\frac{(f_1-f_3)}{2}\mathrm{sech}\Big[\frac{f_1-f_3}{2}\Big(x+\frac{f_1+f_3}{2}t\Big)\Big]+\frac{f_1+f_3}{2}.
\end{equation}

\bigskip

\noindent \textbf{1.b}\quad \textbf{dn solution}\\
Let $y=\mathrm{dn}(\beta\xi)$ with $\xi=x-ct$ where the function satisfies the differential equation \eqref{eq:jacobidnu}. If we compare the coefficients of (\ref{exactsolncaseone}) and \eqref{eq:jacobidnu}, we get
$$
\displaystyle \beta^2=\alpha^2=\frac{2\mu_0}{-\mu_2+\sqrt{4\mu_0+\mu_2^2}}, \quad
k^2=2+\frac{\mu_2^2}{2\mu_0}-\frac{\mu_2}{2\mu_0}\sqrt{4\mu_0+\mu_2^2}.
$$
Explicitly we have
\begin{equation}\label{parameterscase1ba}
\beta^2=\alpha^2=(f_1-f_3)^2/4, \quad k^2=4(f_1-f_2)(f_2-f_3)/(f_1-f_3)^2,
\end{equation}
or
\begin{equation}\label{parameterscase1bb}
\beta^2=\alpha^2=(f_1+f_3-2f_2)^2/4, \quad k^2=-4(f_1-f_2)(f_2-f_3)/(f_1+f_3-2f_2)^2.
\end{equation}
Hence the solution is
\begin{equation}\label{case1bfirstsoln}
\displaystyle u(x,t)=\pm \frac{(f_1-f_3)}{2} \mathrm{dn}\Big[\frac{f_1-f_3}{2}\Big(x+\frac{f_1+f_3}{2}t\Big)\Big]+\frac{f_1+f_3}{2},
\end{equation}
or
\begin{equation}\label{case1bsecondsoln}
\displaystyle u(x,t)=\pm \frac{(f_1+f_3-2f_2)}{2} \mathrm{dn}\Big[\frac{f_1+f_3-2f_2}{2}\Big(x+\frac{f_1+f_3}{2}t\Big)\Big]+\frac{f_1+f_3}{2}.
\end{equation}
\noindent Let us analyze the limiting points for the parameters (\ref{parameterscase1ba}) and the solution (\ref{case1bfirstsoln}). Similar analysis can be done for (\ref{parameterscase1bb}) and (\ref{case1bsecondsoln}). For $k=0$, we have either $f_1=f_2$ or $f_2=f_3$. But we noted before that we do not have real solutions for these cases. For $k=1$, from (\ref{parameterscase1ba}) we get the relation $2f_2=f_1+f_3$. Thus the corresponding solution is
\begin{equation}\label{case1bsoln}
\displaystyle u(x,t)=\pm \frac{(f_1-f_3)}{2} \mathrm{sech}\Big[\frac{f_1-f_3}{2}\Big(x+\frac{f_1+f_3}{2}t\Big)\Big]+\frac{f_1+f_3}{2}.
\end{equation}

\bigskip

\noindent \textbf{Case 2.}\quad \textbf{Solutions of the form $\displaystyle u(x,t)=f(\xi)=a_1/(a_2+b_2y(\beta\xi))$}

Here we shall find solutions of (\ref{l=2}) having the form $\displaystyle u(x,t)=f(\xi)=a_1/(a_2+b_2y(\beta\xi))$, where $a_1, a_2, b_2, \beta$ are constants and $\xi=x-ct$.
If we use this form in the equation (\ref{F(f)}) we get the following equation:
\begin{eqnarray}\label{exactsolncasetwo}
\displaystyle (y')^2&=& \frac{8d_3b_2^2}{\beta^2a_1^2}y^4+\frac{8}{\beta^2a_1^2}(d_2a_1b_2+4d_3a_2b_2)y^3\nonumber\\
&&+\frac{4}{\beta^2a_1^2}(12d_3a_2^2+a_1^2d_1+6d_2a_1a_2-a_1^2c^2)y^2\nonumber\\
&&+\frac{4}{\beta^2a_1^2b_2}(2a_1^2d_1a_2-ca_1^3-2a_1^2c^2a_2+6d_2a_1a_2^2+8d_3a_2^3)y\nonumber\\
&&+\frac{1}{\beta^2a_1^2b_2^2}(8d_3a_2^4+8d_2a_1a_2^3+4a_1^2d_1a_2^2-4a_1^2c^2a_2^2-4ca_1^3a_2-a_1^4),\nonumber\\
&&
\end{eqnarray}
where $a_1, b_2, \beta\neq 0$. As we did in the previous case we shall again use Jacobi elliptic functions (\ref{eq:jacobisnu})-(\ref{eq:jacobioneoversnu}) and study the special cases for $k=0$ and $k=1$. The differential equations satisfied by these elliptic functions do not have terms
with $y^3$ and $y$. Hence the coefficients of $y^3$ and $y$ should be zero in (\ref{exactsolncasetwo}). Let also $\displaystyle a=a_2/a_1$ and $\displaystyle b=b_2/a_1$, $a_1\neq 0$. Then we get
\begin{eqnarray*}
\displaystyle d_1&=&\frac{c}{2a}+c^2+8a^2d_3\\
&=&\frac{1}{16a}[a(f_1^2+f_2^2+f_3^2+f_4^2)+2a(f_1f_2+f_1f_3+f_1f_4+f_2f_3+f_2f_4+f_3f_4)\\
&&-2(f_1+f_2+f_3+f_4)-16a^3f_1f_2f_3f_4]\\
d_2&=&-4d_3a=(f_1f_2f_4+f_1f_2f_3+f_2f_3f_4+f_1f_3f_4)/8\\
a&=&(f_1f_2f_4+f_1f_2f_3+f_2f_3f_4+f_1f_3f_4)/4f_1f_2f_3f_4
\end{eqnarray*}
with a relation between the zeros of $F(f)$:
\begin{multline}\label{case2relationonzeros}
(f_1f_2f_3-f_2f_3f_4-f_1f_2f_4+f_1f_3f_4)(f_1f_2f_3-f_2f_3f_4+f_1f_2f_4-f_1f_3f_4)\\
\times(f_1f_2f_3+f_2f_3f_4-f_1f_2f_4-f_1f_3f_4)=0.
\end{multline}
Hence \eqref{exactsolncasetwo} is simplified as
\begin{equation}\label{simplifiedexactsolncasetwo}
\displaystyle (y')^2=\frac{1}{\beta^2}\nu_4y^4+\frac{1}{\beta^2}\nu_2y^2+\frac{1}{\beta^2b^2}\nu_0,
\end{equation}
where
\begin{eqnarray*}
\nu_4&=&-b^2f_1f_2f_3f_4\\
\nu_2&=&\frac{(f_1f_2f_3+f_1f_2f_4+f_1f_3f_4+f_2f_3f_4)^2}{8f_1f_2f_3f_4}-\frac{2f_1f_2f_3f_4(f_1+f_2+f_3+f_4)}{f_1f_2f_3+f_1f_2f_4+f_1f_3f_4+f_2f_3f_4}\\
\nu_0&=&\frac{(f_1f_2f_3+f_1f_2f_4+f_1f_3f_4+f_2f_3f_4)(f_1+f_2+f_3+f_4)}{8f_1f_2f_3f_4}\\
&&-\frac{(f_1f_2f_3+f_1f_2f_4+f_1f_3f_4+f_2f_3f_4)^4}{256(f_1f_2f_3f_4)^3}-1,
\end{eqnarray*}
for $f_1f_2f_3f_4\neq 0$. If any one of the roots of $F$ is zero i.e. $f_1f_2f_3f_4=0$ then (\ref{case2relationonzeros}) implies that one more
root is also zero. Hence in such a case $F$ has a double zero and two
simple zeros. This case was studied in section 3.1.

Now let us study the elliptic functions satisfying (\ref{simplifiedexactsolncasetwo}). Note that if $f_1f_2f_3f_4\neq 0$, we will take $\displaystyle f_4=f_1f_2f_3/(f_2f_3+f_1f_2-f_1f_3)$ by the relation (\ref{case2relationonzeros}) in the below computations.

\noindent \textbf{2.a}\quad \textbf{sn solution}\\
Let $y=\mathrm{sn}(\beta\xi)$ with $\xi=x-ct$ where the function $y$ satisfies the first order differential equation \eqref{eq:jacobisnu}. Then when we compare the coefficients of (\ref{simplifiedexactsolncasetwo}) and \eqref{eq:jacobisnu},
we get
\begin{equation}\label{parameterscase2a}
\displaystyle \beta^2=-\nu_4-\nu_2,\quad k^2=-\frac{\nu_4}{\nu_4+\nu_2}, \quad b^2=-\frac{\nu_0}{\nu_4+\nu_2}.
\end{equation}
Explicitly, we have
\begin{eqnarray}\label{case2asnsystem}
\displaystyle \beta^2&=&\frac{2b^2f_1^2f_2^2f_3^2-f_1^2f_2^2+2f_1^2f_2f_3-2f_1^2f_3^2+2f_2f_3^2f_1-f_2^2f_3^2}{f_2f_3+f_1f_2-f_1f_3} \nonumber\\
k^2&=&\frac{2b^2f_1^2f_2^2f_3^2}{2b^2f_1^2f_2^2f_3^2-f_1^2f_2^2+2f_1^2f_2f_3-2f_1^2f_3^2+2f_2f_3^2f_1-f_2^2f_3^2},
\end{eqnarray}
and we obtain four choices for the value $b$; $\displaystyle \pm (f_1-f_3)/2f_1f_3$, $\displaystyle \pm (f_1f_2-2f_1f_3+f_2f_3)/2f_1f_2f_3$.
Taking $\displaystyle b=(f_1-f_3)/2f_1f_3$ yields
\begin{eqnarray}\label{case2Aparameterswithzeros}
\beta^2&=&-\frac{f_2^2(f_1+f_3)^2-4f_1f_3(f_1f_2+f_2f_3-f_1f_3)}{4(f_1f_2+f_2f_3-f_1f_3)}\nonumber \\
k^2&=&\frac{f_2^2(f_1-f_3)^2}{f_2^2(f_1+f_3)^2-4f_1f_3(f_1f_2+f_2f_3-f_1f_3)}.
\end{eqnarray}
Hence the solution is
\begin{equation}
\displaystyle u(x,t)=\frac{2f_1f_3}{(f_1+f_3)+(f_1-f_3)\mathrm{sn}[\beta(x-ct)]},
\end{equation}
where
$$
c=-\frac{f_1+f_2+f_3}{4}-\frac{f_1f_2f_3}{4(f_2f_3+f_1f_2-f_1f_3)}.
$$

Let us study the limiting cases. For $k=0$, there are two possibilities: $f_2=0$ or $f_1=f_3$. If $f_2=0$ then $\beta=\pm \sqrt{f_1f_3}$, $f_1f_3> 0$
and the corresponding solution is
\begin{equation}
\displaystyle u(x,t)=\frac{2f_1f_3}{(f_1+f_3)\pm (f_1-f_3)\sin\Big[\sqrt{f_1f_3}(x+\frac{f_1+f_3}{4}t)\Big]}.
\end{equation}
If $f_1=f_3$ then $\displaystyle a=1/f_1$ and $b=0$ so we have constant solution $u(x,t)=f_1$. For $k=1$ then from (\ref{case2Aparameterswithzeros}) we have
$$
4f_1f_3(f_2-f_1)(f_2-f_3)=0.
$$
It is not possible to have $f_1=0$ or $f_3=0$ because of the definition of $b$. If $f_1=f_2$ or
$f_2=f_3$ we have $\beta^2\leq 0$. Hence we do not have real solution for $k=1$.

\bigskip

\noindent \textbf{2.b}\quad \textbf{cn solution}\\
Let $y=\mathrm{cn}(\beta\xi)$ with $\xi=x-ct$ where the function $y$ satisfies the first order differential equation \eqref{eq:jacobicnu}. If we compare the coefficients of (\ref{simplifiedexactsolncasetwo}) and \eqref{eq:jacobicnu},
we get
\begin{eqnarray}\label{parameterscase2b}
\displaystyle \beta^2=-2\nu_4-\nu_2, \quad
k^2=\frac{\nu_4}{2\nu_4+\nu_2}, \quad
b^2=-\frac{\nu_0}{\nu_4+\nu_2}.
\end{eqnarray}
Since we have the same relation for $b$ as in the Case $2.a$, we may also take $\displaystyle b=(f_1-f_3)/2f_1f_3$. Hence (\ref{parameterscase2b}) becomes
\begin{equation}\label{case2bcnsystem}
\displaystyle \beta^2=\frac{f_1f_3(f_1-f_2)(f_2-f_3)}{f_2f_3+f_1f_2-f_1f_3}, \quad
k^2=\frac{f_2^2(f_1-f_3)^2}{4f_1f_3(f_1-f_2)(f_2-f_3)}.
\end{equation}
Thus the solution is
\begin{equation}
\displaystyle u(x,t)=\frac{2f_1f_3}{(f_1+f_3)+(f_1-f_3)\mathrm{cn}[\beta(x-ct)]},
\end{equation}
where
$$
c=-\frac{f_1+f_2+f_3}{4}-\frac{f_1f_2f_3}{4(f_2f_3+f_1f_2-f_1f_3)}.
$$
For $k=0$, there are two possibilities: $f_2=0$ or $f_1=f_3$. If $f_2=0$ then $\beta=\pm \sqrt{f_1f_3}$, $f_1f_3>0$
and the corresponding solution is
\begin{equation}
\displaystyle u(x,t)=\frac{2f_1f_3}{(f_1+f_3)+(f_1-f_3)\cos[\sqrt{f_1f_3}(x+\frac{f_1+f_3}{4}t)]}.
\end{equation}
If $f_1=f_3$ then $\displaystyle a=1/f_1$ and $b=0$ so we have a constant solution $u(x,t)=f_1$. For $k=1$, we have the following relation
from (\ref{case2bcnsystem}):
\begin{equation}
2f_1f_3=f_2(f_1+f_3).
\end{equation}
Hence the solution is
\begin{equation}\label{case2bsoln}
\displaystyle u(x,t)=\frac{2f_1f_3}{(f_1+f_3)+(f_1-f_3)\mathrm{sech}\Big[\frac{f_1-f_3}{f_1+f_3}\sqrt{f_1f_3}(x-ct)\Big]},
\end{equation}
where $\displaystyle c=-(f_1^2+6f_1f_3+f_3^2)/4(f_1+f_3)$.
\bigskip

\noindent \textbf{2.c}\quad \textbf{dn solution}\\
Let $y=\mathrm{dn}(\beta\xi)$ with $\xi=x-ct$ where the function $y$ satisfies the first order differential equation \eqref{eq:jacobidnu}. When we compare the coefficients of (\ref{simplifiedexactsolncasetwo}) and \eqref{eq:jacobidnu},
we get
\begin{equation}\label{parameterscase2c}
\displaystyle \beta^2=-\nu_4 , \quad
k^2=\frac{2\nu_4+\nu_2}{\nu_4} , \quad
b^2=-\frac{\nu_0}{\nu_4+\nu_2}.
\end{equation}
 Same as before let us take $\displaystyle b=(f_1-f_3)/2f_1f_3$. Hence (\ref{parameterscase2c}) becomes
\begin{equation}\label{case2cdnsystem}
\displaystyle \beta^2=\frac{f_2^2(f_1-f_3)^2}{4(f_2f_3+f_1f_2-f_1f_3)}, \quad
k^2=\frac{4f_1f_3(f_1-f_2)(f_2-f_3)}{f_2^2(f_1-f_3)^2}.
\end{equation}
Thus the solution is
\begin{equation}
\displaystyle u(x,t)=\frac{2f_1f_3}{(f_1+f_3)+(f_1-f_3)\mathrm{dn}[\beta(x-ct)]},
\end{equation}
where
$$
\displaystyle c=-\frac{f_1+f_2+f_3}{4}-\frac{f_1f_2f_3}{4(f_2f_3+f_1f_2-f_1f_3)}.
$$
For $k=0$, there are four possibilities: $f_1=0$, $f_3=0$, $f_1=f_2$ or $f_2=f_3$. We cannot have $f_1=0$ or $f_3=0$ because of the
definition of $b$. If $f_1=f_2$ or $f_2=f_3$, the solution is $u(x,t)=f_3$.
For $k=1$, we have $2f_1f_3=f_2(f_1+f_3)$. So the corresponding solution is
\begin{equation}\label{case2csoln}
\displaystyle u(x,t)=\frac{2f_1f_3}{(f_1+f_3)+(f_1-f_3)\mathrm{sech}\Big[\frac{f_1-f_3}{f_1+f_3}\sqrt{f_1f_3}(x-ct)\Big]}, \quad f_1f_3>0,
\end{equation}
where $\displaystyle c=-(f_1^2+6f_1f_3+f_3^2)/4(f_1+f_3)$.
\bigskip

\noindent \textbf{2.d}\quad \textbf{tn solution}\\
Let $y=\mathrm{tn}(\beta\xi)$ with $\xi=x-ct$ where the function $y$ satisfies the first order differential equation \eqref{eq:jacobitnu}. Hence when we compare the coefficients of (\ref{simplifiedexactsolncasetwo}) and \eqref{eq:jacobitnu},
we get
\begin{equation}\label{parameterscase2d}
\displaystyle \beta^2=\nu_2-\nu_4, \quad
k^2=\frac{\nu_2-2\nu_4}{\nu_2-\nu_4}, \quad
b^2=\frac{\nu_0}{\nu_2-\nu_4}.
\end{equation}
Here we notice that
third equality of (\ref{parameterscase2d}) reveals that $b$ is not real for any values of $k$.
Hence for all values of $k^2\in [0,1]$ we do not have real solution.

\bigskip

\noindent \textbf{2.e}\quad  \textbf{1/sn solution}\\
Let $\displaystyle y=1/\mathrm{sn}(\beta\xi)$ with $\xi=x-ct$ where the function $y$ satisfies the first order differential equation \eqref{eq:jacobioneoversnu}. Hence when we compare the coefficients of (\ref{simplifiedexactsolncasetwo}) and \eqref{eq:jacobioneoversnu},
we get
\begin{equation}\label{parameterscase2e}
\displaystyle \beta^2=\nu_4, \quad
k^2=\frac{-\nu_2-\nu_4}{\nu_4}, \quad
b^2=\frac{-\nu_0}{\nu_2+\nu_4}.
\end{equation}
If we take $\displaystyle b=(f_1-f_3)/2f_1f_3$,
(\ref{parameterscase2e}) becomes
\begin{equation}\label{case2Eparameterswithzeros}
\beta^2=-\frac{f_2^2(f_1-f_3)^2}{4(f_2f_3+f_1f_2-f_1f_3)}, \quad
k=\pm\frac{f_2(f_1+f_3)-2f_1f_3}{f_2(f_1-f_3)}.
\end{equation}
The corresponding solution is
\begin{equation}
\displaystyle u(x,t)=\frac{2f_1f_3\mathrm{sn}[\beta(x-ct)]}{(f_1+f_3)\mathrm{sn}[\beta(x-ct)]+(f_1-f_3)},
\end{equation}
where
$$
c=-\frac{f_1+f_2+f_3}{4}-\frac{f_1f_2f_3}{4(f_2f_3+f_1f_2-f_1f_3)}.
$$
For $k=0$, we have the relation $f_2(f_1+f_3)=2f_1f_3$ and the solution becomes
\begin{equation}
\displaystyle u(x,t)=\frac{2f_1f_3\sin\Big[\frac{f_1-f_3}{f_1+f_3}
\sqrt{-f_1f_3}(x-ct)\Big]}{(f_1+f_3)\sin\Big[\frac{f_1-f_3}{f_1+f_3}\sqrt{-f_1f_3}(x-ct)\Big]\pm(f_1-f_3)},\quad -f_1f_3>0,
\end{equation}
where $\displaystyle c=-(f_1^2+6f_1f_3+f_3^2)/4(f_1+f_3)$. For $k=1$ then from (\ref{case2Eparameterswithzeros}) we have
$$
4f_1f_3(f_2-f_1)(f_2-f_3)=0.
$$
It is not possible to have $f_1=0$ or $f_3=0$ because of the definition of $b$. If $f_1=f_2$ or
$f_2=f_3$ we have $\beta^2< 0$. Hence we do not have real solution for $k=1$.

\bigskip

\noindent \textbf{2.f}\quad \textbf{1/cn solution}\\
Let $\displaystyle y=1/\mathrm{cn}(\beta\xi)$ with $\xi=x-ct$ where the function $y$ satisfies the first order differential equation \eqref{eq:jacobioneovercnu}. If we compare the coefficients of (\ref{simplifiedexactsolncasetwo}) and \eqref{eq:jacobioneovercnu},
we get
\begin{equation}\label{parameterscase2f}
\displaystyle \beta^2=\nu_2+2\nu_4, \quad
k^2=\frac{\nu_2+\nu_4}{\nu_2+2\nu_4}, \quad
b^2=\frac{-\nu_0}{\nu_2+\nu_4}.
\end{equation}
Since we take $\displaystyle b=(f_1-f_3)/2f_1f_3$,
(\ref{parameterscase2f}) becomes
$$
\beta^2=-\frac{f_1f_3(f_1-f_2)(f_2-f_3)}{f_2f_3+f_1f_2-f_1f_3}, \quad
k^2=-\frac{[f_2(f_1+f_3)-2f_1f_3]^2}{4f_1f_3(f_1-f_2)(f_2-f_3)}.
$$
The corresponding solution is
\begin{equation}
\displaystyle u(x,t)=\frac{2f_1f_3\mathrm{cn}[\beta(x-ct)]}{(f_1+f_3)\mathrm{cn}[\beta(x-ct)]+(f_1-f_3)},
\end{equation}
where
$$
c=-\frac{f_1+f_2+f_3}{4}-\frac{f_1f_2f_3}{4(f_2f_3+f_1f_2-f_1f_3)}.
$$
For $k=0$, we have the relation $f_2(f_1+f_3)=2f_1f_3$ and the solution becomes
\begin{equation}
\displaystyle u(x,t)=\frac{2f_1f_3\cos\Big[\frac{f_1-f_3}{f_1+f_3}
\sqrt{-f_1f_3}(x-ct)\Big]}{(f_1+f_3)\cos\Big[\frac{f_1-f_3}{f_1+f_3}\sqrt{-f_1f_3}(x-ct)\Big]+(f_1-f_3)},\quad -f_1f_3>0,
\end{equation}
where $\displaystyle c=-(f_1^2+6f_1f_3+f_3^2)/4(f_1+f_3)$. The case for $k=1$ gives the condition $f_2^2(f_1-f_3)^2=0$ to be satisfied.
Hence we have two possibilities: $f_2=0$ or $f_1=f_3$.
If $f_2=0$ then $\beta=\pm \sqrt{-f_1f_3}$, $-f_1f_3> 0$
and the corresponding solution is
\begin{equation}\label{case2fsoln}
\displaystyle u(x,t)=\frac{2f_1f_3\mathrm{sech}[\sqrt{-f_1f_3}(x+\frac{f_1+f_3}{4}t)]}{(f_1+f_3)\mathrm{sech}[\sqrt{-f_1f_3}(x+\frac{f_1+f_3}{4}t)]+(f_1-f_3)},\quad -f_1f_3>0.
\end{equation}
If $f_1=f_3$ then $\displaystyle a=1/f_1$ and $b=0$ so we have a constant solution $u(x,t)=f_1$.

\bigskip

\noindent \textbf{2.g} \textbf{dn\,tn solution}\\
Let $y=\mathrm{dn}(\beta\xi)\, \mathrm{tn}(\beta\xi)$ with $\xi=x-ct$ where the function $y$ satisfies the first order differential equation \eqref{eq:jacobidnutnu}. Hence when we compare the coefficients of (\ref{simplifiedexactsolncasetwo}) and \eqref{eq:jacobidnutnu},
we get
$$
\displaystyle \beta^2=\nu_4, \quad
k^2=\frac{2\nu_4-\nu_2}{4\nu_4}, \quad
b^2=\frac{\nu_0}{\nu_4}.
$$
The third equality above gives four choices for
$b$:
\begin{equation}
\displaystyle \pm \frac{\sqrt{-f_2(f_3-f_1)(2f_1f_3-f_2(f_1+f_3))}}{f_1f_2f_3} \quad  \quad \pm \frac{\sqrt{f_2(f_3-f_1)(2f_1f_3-f_2(f_1+f_3))}}{f_1f_2f_3}.
\end{equation}
To have real solutions, the parameters must be real. Hence from the expressions for $b$ we have either
$-f_2(f_3-f_1)(2f_1f_3-f_2(f_1+f_3))\geq 0$ or $f_2(f_3-f_1)(2f_1f_3-f_2(f_1+f_3))\geq 0$. If the first one is true then
\begin{equation}\label{firstgroup2g}
\quad \beta^2=\frac{f_2(f_3-f_1)(2f_1f_3-f_2(f_1+f_3))}{4(f_2f_3+f_1f_2-f_1f_3)}, \quad
k^2=\frac{-f_3^2(f_1-f_2)^2}{f_2(f_3-f_1)(2f_1f_3-f_2(f_1+f_3))}.
\end{equation}
If the second one is true then
\begin{equation}\label{secondgroup2g}
\quad \beta^2=\frac{f_2(f_1-f_3)(2f_1f_3-f_2(f_1+f_3))}{4(f_2f_3+f_1f_2-f_1f_3)}, \quad
k^2=\frac{f_1^2(f_2-f_3)^2}{f_2(f_1-f_3)(-2f_1f_3+f_2(f_1+f_3))}.
\end{equation}
From the equality for $k^2$ in (\ref{firstgroup2g}) we get
$$
\displaystyle \frac{k^2-1}{k^2}=\frac{f_1^2(f_3-f_2)^2}{f_3^2(f_2-f_1)^2}\geq 0.
$$
This gives that $k^2\geq 1$. We know that for the parameter $k^2$ of Jacobi elliptic functions we have $0\leq k^2\leq 1$. Additionally, at the limiting points
$k=0$ and $k=1$ it yields that $F(f)$ has two double zeros that is the case which does not give real solution as we stated in section $2.2$. We also have the similar result for
(\ref{secondgroup2g}). Hence we do not have real solutions for all $k^2\in [0,1]$.

\bigskip

\subsection{Discussion About the Special Solutions}

When $F(f)$ has one double $f_2$ and two simple zeros $f_1$ and $f_3$ we have the following system of equations:
\begin{eqnarray}\label{systemofeqn}
-4c&=&f_1+2f_2+f_3\nonumber\\
4(d_1-c^2)&=&-[f_1f_3+2f_2(f_1+f_3)+f_2^2]\nonumber \\
8d_2&=&2f_1f_2f_3+f_2^2(f_1+f_3)\nonumber \\
8d_3&=&-f_1f_2^2f_3.
\end{eqnarray}
The exact solutions in terms of the Jacobi elliptic functions take the following forms:

\noindent \textbf{(i)} In Case $1.a$ and Case $1.b$, we have $d_2=cd_1$. Using this in (\ref{systemofeqn}) we obtain that
either $2f_2=f_1+f_3$ or $f_1=f_3$. The second one is not allowed due to the discussion in the section $3.2$. By using
(\ref{systemofeqn}), the first one leads to $k^2=1$. In this case the solution is given in (\ref{case1asoln}) and (\ref{case1bsoln})
which are compatible with the limiting solutions discussed in section $5$, part $a$.

\noindent \textbf{(ii)} In Case $2.b$ and Case $2.c$, we have $d_2=-4d_3a$. From the first equation of (\ref{systemofeqn})
we have $\displaystyle a=1/f_2$. Then this implies $\displaystyle d_2=-4d_3/f_2$. This constraint gives
$2f_1f_3=f_2(f_1+f_3)$ which yields $k^2=1$. In this case the solutions are given in (\ref{case2bsoln}) and (\ref{case2csoln})
which are compatible with the limiting solutions discussed in section $5$, part $b$.

\noindent \textbf{(iii)} If $f_2=0$ then $d_3=0$ hence $d_2=0$ which leads to $k^2=1$. In this case the solution is (\ref{case2fsoln}) given in
Case $2.f$ which are compatible with the limiting solutions discussed in section $5$, part $c$.

\subsection{General Solutions of (\ref{l=2}) in Terms of Elliptic Functions}

Here we shall deal with the most general form of solutions
\begin{equation}\label{generalformofsoln}
\displaystyle u(x,t)=f(\xi)=(a_1+b_1y(\beta\xi))/(a_2+b_2y(\beta\xi)),\quad \xi=x-ct.
\end{equation}
When we insert this form into the equation $F(f)$ we get
\begin{equation}\label{generalsoln}
\displaystyle (y')^2=\frac{1}{\beta^2(b_1a_2-b_2a_1)^2}(\Omega_4y^4+\Omega_3y^3+\Omega_2y^2+\Omega_1y+\Omega_0),
\end{equation}
where
\begin{eqnarray}
\Omega_4&=&-b_1^2(b_1+2cb_2)^2+4b_2^2(b_1^2d_1+2b_1b_2d_2+2b_2^2d_3)=b_2^4F(b),\quad b=b_1/b_2\\
\Omega_3&=&4(2d_2a_1b_2^3-cb_1^3a_2+8d_3a_2b_2^3-3ca_1b_1^2b_2+2d_1a_1b_1b_2^2+2d_1b_1^2a_2b_2\nonumber\\
&&-2c^2a_1b_1b_2^2-2c^2b_1^2a_2b_2+6d_2b_1a_2b_2^2-a_1b_1^3)\\
\Omega_2&=&4d_1a_1^2b_2^2+4d_1b_1^2a_2^2-4c^2a_1^2b_2^2-4c^2b_1^2a_2^2+48d_3a_2^2b_2^2-12ca_1^2b_1b_2-12ca_1b_1^2a_2
\nonumber\\&&+24d_2a_1a_2b_2^2+24d_2b_1a_2^2b_2
-6a_1^2b_1^2+16d_1a_1b_1a_2b_2-16c^2a_1b_1a_2b_2\\
\Omega_1&=&4(2d_2b_1a_2^3-ca_1^3b_2+8d_3a_2^3b_2-3ca_1^2b_1a_2+2d_1a_1^2a_2b_2+2d_1a_1b_1a_2^2\nonumber\\
&&-2c^2a_1^2a_2b_2-2c^2a_1b_1a_2^2+6d_2a_1a_2^2b_2-a_1^3b_1)\\
\Omega_0&=&-a_1^4+8d_3a_2^4-4ca_1^3a_2+4d_1a_1^2a_2^2-4c^2a_1^2a_2^2+8d_2a_1a_2^3=a_2^4F(a), \quad a=a_1/a_2,\nonumber\\
&&
\end{eqnarray}
with $\, b_1a_2-b_2a_1\neq 0$. We have four arbitrary constants $c, d_1, d_2, d_3$ in the differential equation (\ref{F(f)}). In (\ref{generalsoln}) we have
effectively four independent parameters. By choosing these constants properly we get several solutions in terms of elliptic functions. We can analyze these solutions in two groups:

\noindent\textbf{i)} If $F(f)$ has zeros then we can make the coefficients of $y^4$ to vanish by taking $F(b)=0$. This
means that $b=b_1/b_2$ is a zero of $F(f)$. In addition to that choosing the constant $\Omega_0=0$ yields that $F(a)=0$ where $a=a_1/a_2$. This
also means that $a=a_1/a_2$ is another zero of $F(f)$. Note that $a\neq b$ since $b_1a_2-b_2a_1\neq 0$. Then the equation (\ref{generalsoln}) takes
the form where the square of elliptic functions and their inverses given in (\ref{eq:jacobisquaresn})-(\ref{eq:jacobisquaredntn}) satisfy. By making substitution $a=a_1/a_2$ and $b=b_1/b_2$, the equation (\ref{generalsoln}) becomes
\begin{eqnarray}\label{generalsolncasei}
\displaystyle (y')^2&=&\frac{b_2^2F(b)}{\beta^2a_2^2(b-a)^2}y^4+\frac{4b_2}{\beta^2a_2(b-a)^2}\omega_3y^3+\frac{2}{\beta^2(b-a)^2}\omega_2y^2\nonumber\\
&&+\frac{4a_2}{\beta^2b_2(b-a)^2}\omega_1y+\frac{a_2^2F(a)}{\beta^2b_2^2(b-a)^2}
\end{eqnarray}
where
\begin{eqnarray*}
\omega_3&=&2ad_3-cb^3+8d_3-3acb^2+2d_1ab+2d_1b^2-2c^2ab-2c^2b^2+6d_2b-ab^3 \\
\omega_2&=&2d_1a^2+2d_1b^2-2c^2a^2-2c^2b^2+24d_3-6ca^2b-6cab^2+12d_2a+12d_2b
\\&&-3a^2b^2+8d_1ab-8c^2ab\\
\omega_1&=&2d_2b-ca^3+8d_3-3bca^2+2d_1ab+2d_1a^2-2c^2ab-2c^2a^2+6d_2a-ba^3.
\end{eqnarray*}
If $a$ and $b$ are the zeros of $F(f)$, then $F(a)=F(b)=0$ and we do not have the terms with $y^4$ and the constant term in (\ref{generalsolncasei}).
For instance, let $a=f_1$ and $b=f_2$, then we can write $F(f)=-(f-f_1)(f-f_2)(f-f_3)(f-f_4)$ such that $f_1,f_2,f_3,f_4$ are zeros of $F(f)$.
Let us write $\omega_1$, $\omega_2$ and $\omega_3$ in terms of the zeros of the function $F(f)$ by the help of (\ref{parameters}).
\begin{eqnarray}\label{omegasystem}
\displaystyle \omega_1
&=&\frac{(f_1-f_2)^2}{4}(f_1-f_4)(f_1-f_3)\\
\omega_2&=&\frac{1}{2}(f_1-f_2)^2\{(f_2-f_3)(f_1-f_4)+(f_1-f_3)(f_2-f_4)\}\\
\omega_3&=&\frac{(f_1-f_2)^2}{4}(f_2-f_3)(f_2-f_4).
\end{eqnarray}
Now we give all solutions of (\ref{l=2}) of the form (\ref{generalformofsoln}). Let $y=\mathrm{sn}^2(\beta\xi)$
with $\xi=x-ct$ where the function $y$ satisfies the first order differential equation (\ref{eq:jacobisquaresn}). Hence when we compare the coefficients of (\ref{generalsolncasei}) and (\ref{eq:jacobisquaresn}), we get
$$
\displaystyle \beta^4=\frac{\omega_1\omega_3}{k^2(b-a)^4},\quad
\frac{b_2}{a_2}=\frac{-2(1+k^2)\omega_1}{\omega_2}$$
$$k^2=-1+\frac{\omega_2^2}{8\omega_1\omega_3}\pm \frac{\omega_2}{8\omega_1\omega_3}\sqrt{\omega_2^2-16\omega_1\omega_3}.
$$
Here without loosing any generality we take $f_1\leq f_2\leq f_3 \leq f_4$. In the case of equality between the zeros of $F(f)$ we have either
 one double zero and two simple zeros or one triple zero and one simple zero cases to have real solutions. Both of these cases were studied in section 3.1 and section 3.2. Therefore we assume, in the sequel, that we have $f_1< f_2< f_3 < f_4$.

\noindent \textbf{(1)}\, Let \textbf{$a=f_1$, $b=f_2$}. For this choice we have
$$
\displaystyle \beta=\frac{1}{2}\sqrt{(f_1-f_3)(f_2-f_4)}, \quad
\frac{b_2}{a_2}=\frac{f_4-f_1}{f_2-f_4}, \quad
k^2=\frac{(f_2-f_3)(f_1-f_4)}{(f_1-f_3)(f_2-f_4)},
$$
and hence the solution with the initial condition $f(0)=f_1$ is
\begin{equation}
\displaystyle u(x,t)=f(\xi)=\frac{f_1(f_2-f_4)+f_2(f_4-f_1)\mathrm{sn}^2((1/2)\sqrt{(f_1-f_3)(f_2-f_4)}\xi)}
{(f_2-f_4)+(f_4-f_1)\mathrm{sn}^2((1/2)\sqrt{(f_1-f_3)(f_2-f_4)}\xi)}.
\end{equation}
Note that since $a$ and $b$ are any zeros of $F(f)$ we have other choices of these parameters.

\noindent \textbf{(2)}\, Let \textbf{$a=f_2$, $b=f_1$}. For this choice we have
$$
\displaystyle \beta=\frac{1}{2}\sqrt{(f_1-f_3)(f_2-f_4)}, \quad
\frac{b_2}{a_2}=\frac{f_3-f_2}{f_1-f_3}, \quad
k^2=\frac{(f_2-f_4)(f_1-f_3)}{(f_1-f_4)(f_2-f_3)},
$$
and hence the solution with the initial condition $f(0)=f_2$ is
\begin{equation}
\displaystyle u(x,t)=f(\xi)=\frac{f_2(f_1-f_3)+f_1(f_3-f_2)\mathrm{sn}^2((1/2)\sqrt{(f_1-f_3)(f_2-f_4)}\xi)}
{(f_1-f_3)+(f_3-f_2)\mathrm{sn}^2((1/2)\sqrt{(f_1-f_3)(f_2-f_4)}\xi)}.
\end{equation}

\noindent \textbf{(3)}\, Let \textbf{$a=f_3$, $b=f_4$}. For this choice we have
$$
\displaystyle \beta=\frac{1}{2}\sqrt{(f_1-f_3)(f_2-f_4)}, \quad
\frac{b_2}{a_2}=\frac{f_2-f_3}{f_4-f_2}, \quad
k^2=\frac{(f_4-f_1)(f_3-f_2)}{(f_3-f_1)(f_4-f_2)},
$$
and hence the solution with the initial condition $f(0)=f_3$ is
\begin{equation}
\displaystyle u(x,t)=f(\xi)=\frac{f_3(f_4-f_2)+f_4(f_2-f_3)\mathrm{sn}^2((1/2)\sqrt{(f_1-f_3)(f_2-f_4)}\xi)}
{(f_4-f_2)+(f_2-f_3)\mathrm{sn}^2((1/2)\sqrt{(f_1-f_3)(f_2-f_4)}\xi)}.
\end{equation}

\noindent \textbf{(4)}\, Let \textbf{$a=f_4, b=f_3$}. For this choice we have
$$
\displaystyle \beta=\frac{1}{2}\sqrt{(f_1-f_3)(f_2-f_4)}, \quad
\frac{b_2}{a_2}=\frac{f_1-f_4}{f_3-f_1}, \quad
k^2=\frac{(f_3-f_2)(f_4-f_1)}{(f_3-f_1)(f_4-f_2)},
$$
and hence the solution with the initial condition $f(0)=f_4$ is
\begin{equation}
\displaystyle u(x,t)=f(\xi)=\frac{f_4(f_3-f_1)+f_3(f_1-f_4)\mathrm{sn}^2((1/2)\sqrt{(f_1-f_3)(f_2-f_4)}\xi)}
{(f_3-f_1)+(f_1-f_4)\mathrm{sn}^2((1/2)\sqrt{(f_1-f_3)(f_2-f_4)}\xi)}.
\end{equation}

Similarly, we can also find other type of solutions including square of Jacobi elliptic functions and inverses of them. But
they are equivalent because of the relations $\mathrm{cn}^2v+\mathrm{sn}^2v=1$ and $\displaystyle \mathrm{tn}^2v=\mathrm{sn}^2v/(1-\mathrm{sn}^2v)$. Hence the solutions given in $\textbf{(1)}$, $\textbf{(2)}$, $\textbf{(3)}$ and $\textbf{(4)}$ are all the most general solutions of (\ref{l=2}) depending upon
the initial conditions.

\noindent \textbf{ii)} Another choice is taking $a_1, b_1, a_2, b_2 $ so that $\Omega_3=\Omega_1=0$. Then the equation (\ref{generalsoln})
takes the form where elliptic functions and their inverses given in (\ref{eq:jacobisnu})-(\ref{eq:jacobioneovercnu}) satisfy. Note that if we take $b_2=0$ to make $\Omega_1=\Omega_3=0$ we have $c=-a_1/a_2$, $d_2=cd_1$ and the solution becomes $f(\xi)=\gamma+\alpha y$ which we have already studied in section $4.2$, Case $1$. If $a_2=0$ we can use inverse of Jacobi elliptic functions for $y$ and then the case turns to $b_2=0$ case. When we take $b_1=0$, to make $\Omega_1=\Omega_3=0$ we have $d_2=-4d_3a_2/a_1$ and $d_1=8d_3(a_2/a_1)^2+c^2+c/2a_2$ and the solution becomes $\displaystyle f(\xi)=1/(\alpha+\gamma y)$ that is the case we have already studied in section $4.2$, Case $2$.

 In the next section we mention about the system (\ref{eqn1}) when $\ell=3$ and $\ell=4$.

\section{$\ell=3$ and $\ell=4$ Cases}

\noindent \textbf{1)} The degenerate coupled KdV equation for $\ell=3$ is
\begin{eqnarray}\label{l=3}
\displaystyle u_t&=&\frac{3}{2}uu_x+v_x\nonumber \\
v_t&=&vu_x+\frac{1}{2}uv_x+\omega_x\nonumber \\
\omega_t&=&-\frac{1}{4}u_{xxx}+\omega u_x+\frac{1}{2}u\omega_x.
\end{eqnarray}
Here we will show that unlike the case $\ell=2$,
we have real traveling wave solution with asymptotically
vanishing boundary condition in $\ell=3$ case. Let $u(x,t)=f(\xi)$, $v(x,t)=g(\xi)$, and $\omega(x,t)=h(\xi)$, where $\xi=x-ct$. From the first equation of (\ref{l=3}) we have
$$
\displaystyle -cf'=\frac{3}{2}ff'+g',
$$
which gives
$$\displaystyle g(\xi)=-cf-\frac{3}{4}f^2+d_1,$$
where $d_1$ is an integration constant. Using $g(\xi)$ in the second equation of \eqref{l=3} yields
$$
\displaystyle h'=3cff'+\frac{3}{2}f^2f'+(c^2-d_1)f'.
$$
Integrating above equation once we get
$$
\displaystyle h(\xi)=\frac{3}{2}cf^2+\frac{1}{2}f^3+(c^2-d_1)f+d_2,
$$
where $d_2$ is an integration constant. Using $h(\xi)$ in the third equation of \eqref{l=3} yields
$$
\displaystyle \frac{1}{4}f'''=\Big(\frac{9c^2}{2}-\frac{3d_1}{2}\Big)ff'+\frac{9c}{2}f^2f'+\frac{5}{4}f^3f'+(c^3-cd_1+d_2)f'.
$$
Integrating above equation once we obtain
$$
\displaystyle \frac{1}{4}f''=\Big(\frac{9c^2}{4}-\frac{3d_1}{4}\Big)f^2+\frac{3c}{2}f^3+\frac{5}{16}f^4+(c^3-cd_1+d_2)f+d_3.
$$
By using $f'$ as an integrating factor, we integrate once more. Finally, we get
$$
\displaystyle (f')^2=\frac{f^5}{2}+3cf^4+(6c^2-2d_1)f^3+4(c^3-cd_1+d_2)f^2+8d_3f+8d_4,
$$
where $c, d_1, d_2, d_3, d_4$ are constants. If we apply the boundary conditions  $f,f',f'',f''',g,g',h,h'\rightarrow 0$ as $\xi\rightarrow \pm \infty$ we get $d_1=d_2=d_3=d_4=0$. Hence we have
\begin{eqnarray}
(f')^2&=&\frac{f^5}{2}+3cf^4+6c^2f^3+4c^3f^2\nonumber\\
&=&\frac{f^2}{2}(f+2c)^3.
\end{eqnarray}
By using trigonometric substitution $f=-2c\sin^2\theta$ and making the cancelations, above equality becomes
$$
\displaystyle \frac{d\theta}{c^{3/2}\sin\theta \cos^2\theta}=\mp d\xi \quad \Rightarrow \frac{\sin\theta d\theta}{c^{3/2}\sin^2\theta \cos^2\theta}=\mp d\xi.
$$
Making the substitution $u=\cos{\theta}$ gives
$$
\displaystyle \frac{du}{c^{3/2}(u^2-1)u^2}=\mp d\xi,
$$
which is solved as
\begin{equation}
\displaystyle \frac{1}{c^{3/2}}\Big\{\frac{1}{u}+\frac{1}{2}\ln|u-1|-\frac{1}{2}\ln|u+1|\Big\}=\mp(\xi-{\xi}_0),
\end{equation}
where ${\xi}_0$ is an integration constant. Note that $\displaystyle u=\cos{\theta}=\pm \Big(1+\frac{f}{2c}\Big)^{1/2}$.
When the solution $f=0$, $u$ is either $1$ or $-1$. Insert the expression for $u$ into the above equation so we get the relation
defining the solution $f$,
\begin{equation}
\displaystyle \frac{1}{c^{3/2}}\left\{\pm \Big(1+\frac{f}{2c}\Big)^{-1/2}+\ln\left|\frac{\pm \Big(1+\frac{f}{2c}\Big)^{1/2}-1}{\pm \Big(1+\frac{f}{2c}\Big)^{1/2}+1}\right|\right\}=\mp(\xi-{\xi}_0).
\end{equation}
Hence we have asymptotically vanishing real traveling solution for $\ell=3$. We expect that this is true for all odd $\ell$.
\bigskip

\noindent \textbf{2)} Now let us analyze $\ell=4$ case. The degenerate coupled KdV equation for $\ell=4$ is
\begin{eqnarray}\label{l=4}
\displaystyle u_t&=&\frac{3}{2}uu_x+v_x\nonumber\\
v_t&=&vu_x+\frac{1}{2}uv_x+\omega_x\nonumber\\
\omega_t&=&\omega u_x+\frac{1}{2}u\omega_x+\rho_x\nonumber\\
\rho_t&=&-\frac{1}{4}u_{xxx}+\rho u_x+\frac{1}{2}u\rho_x.
\end{eqnarray}

\begin{pro} There is no real asymptotically vanishing traveling wave solution of the equation (\ref{l=4}) in the
form $u(x,t)=f(\xi)$, $v(x,t)=g(\xi)$, $\omega(x,t)=h(\xi)$ and  $\rho(x,t)=r(\xi)$, where $\xi=x-ct$.
\end{pro}

\noindent \textbf{Proof.} \, Let $u(x,t)=f(\xi)$, $v(x,t)=g(\xi)$, $\omega(x,t)=h(\xi)$ and $\rho(x,t)=r(\xi)$, where $\xi=x-ct$. From the first equation of (\ref{l=4}) we have
$$
\displaystyle -cf'=\frac{3}{2}ff'+g',
$$
which gives
$$\displaystyle g(\xi)=-cf-\frac{3}{4}f^2+d_1,$$
where $d_1$ is an integration constant. Using $g(\xi)$ in the second equation of \eqref{l=4} yields
$$
\displaystyle h'=3cff'+\frac{3}{2}f^2f'+(c^2-d_1)f'.
$$
Integrating above equation once we have
$$
\displaystyle h(\xi)=\frac{3}{2}cf^2+\frac{1}{2}f^3+(c^2-d_1)f+d_2,
$$
where $d_2$ is an integration constant. Using $h(\xi)$ in the third equation of \eqref{l=4} yields
$$
\displaystyle r'=-\frac{5}{4}f^3f'-\frac{9}{2}cf^2f'-\Big(\frac{9}{2}c^2+\frac{3}{2}d_1\Big)ff'+(-c^3+cd_1-d_2)f'.
$$
Integrating this equation once gives
$$
\displaystyle r(\xi)=-\frac{5}{16}f^4-\frac{3}{2}cf^3+\Big(-\frac{9}{4}c^2+\frac{3}{4}d_1\Big)f^2+(-c^3+cd_1-d_2)f+d_3,
$$
where $d_3$ is an integration constant. Using $r(\xi)$ in the fourth equation of \eqref{l=4} gives
\begin{eqnarray*}
\displaystyle \frac{1}{4}f'''&=&-\frac{15}{16}f^4f'-5cf^3f'+\Big(\frac{3}{2}d_1-9c^2\Big)f^2f'+\Big(3cd_1-6c^3-\frac{3}{2}d_2\Big)ff'
\\&&+(c^2d_1+d_3-c^4-cd_2)f'.
\end{eqnarray*}
Integrating the above equation once we get
\begin{eqnarray*}
\displaystyle \frac{1}{4}f''&=&-\frac{3}{16}f^5-\frac{5c}{4}f^4+\Big(\frac{d_1}{2}-3c^2\Big)f^3+\Big(\frac{3}{2}cd_1-3c^3-\frac{3}{4}d_2\Big)f^2
\\&&+(c^2d_1+d_3-c^4-cd_2)f+d_4,
\end{eqnarray*}
where $d_4$ is an integration constant. By using $f'$ as an integrating factor, we integrate once more. Finally, we get
\begin{eqnarray*}
\displaystyle (f')^2&=&-\frac{1}{4}f^6-2cf^5+(d_1-6c^2)f^4+(4cd_1-8c^3-2d_2)f^3
\\&&+4(c^2d_1+d_3-c^4-cd_2)f^2+8d_4f+8d_5,
\end{eqnarray*}
where $d_5$ is an integration constant. If we apply the boundary conditions $f$, $f'$, $f''$, $f'''$, $g$, $g'$, $h$, $h'$, $r$, $r'$ $\rightarrow$ $0$ as $\xi\rightarrow \infty$, we get
$d_1=d_2=d_3=d_4=d_5=0$. Hence the above equation becomes
\begin{eqnarray*}
(f')^2&=&-\frac{1}{4}f^6-2cf^5-6c^2f^4-8c^3f^3-4c^4f^2\\
&=&-\frac{f^2}{4}(f+2c)^4.
\end{eqnarray*}
Obviously, there is no real traveling wave solution of the case $\ell=4$ with asymptotically vanishing boundary conditions.

\noindent \textbf{Conjecture:} For all even $\ell$, since we have the following equality
\begin{equation}
(f')^2=-\frac{f^2}{2^{\ell-2}}(f+2c)^\ell, \quad u(x,t)=f(\xi) \quad \xi=x-ct,
\end{equation}
the degenerate coupled KdV equation (\ref{eqn1}) does not have real traveling wave solution with asymptotically vanishing boundary conditions.

\section{Graphs of the Exact Solutions}
Here we give the graphs of exact solutions to see the behavior of the solutions.

\noindent \textbf{Case $1.a$ and Case $1.b$ for $k=1$:}\\
According to the conditions on parameters, the parameters are chosen as
$$
\alpha=\beta=1, \quad c=2, \quad d_1=-7/4 , \quad d_2=-7/2, \quad d_3=-3/2.
$$
Hence the solution becomes
\begin{equation}
u(x,t)=\mathrm{sech}(\xi)-2, \quad \xi=x-2t,
\end{equation}
and the graph of this function is
\begin{figure}[!h]
\centering
\begin{center}
\includegraphics[angle=0,scale=.30]{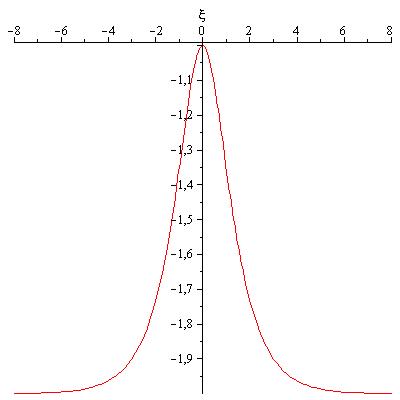}
\end{center}
\caption{Graph of Case $1.a$-$1.b$\, ($k=1$)}
\end{figure}

\newpage
\noindent Note that by the choice of the parameters of this case the equation (\ref{F(f)}) becomes
$$
F(f)=-(f+3)(f+1)(f+2)^2.
$$
The numerical values of the zeros of $F(f)$ are such that the graph corresponds to the exact solitary wave solution given in
section $3.3$, part $(a)$.

\noindent \textbf{Case $1.a$ for different values of $k$:}\\
Here to see the behavior of the solution by the change of the value of $k$ we give the following graph:
\begin{figure}[!h]
\centering
\begin{center}
\includegraphics[angle=0,scale=.30]{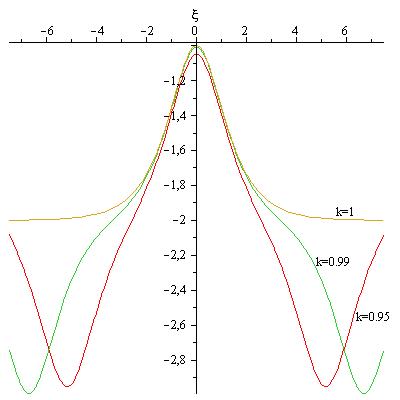}
\end{center}
\caption{Graph of Case $1.a$\, (different values of $k$)}
\end{figure}

\bigskip

\noindent \textbf{Case $1.b$ for $k=0.5$:}\\
The parameters are chosen as
$$
\displaystyle \alpha=\beta=1, \quad c=2 , \quad d_1=-\frac{25}{16}
, \quad d_2=-\frac{25}{8} , \quad d_3=-\frac{39}{32}.
$$
The solution is
\begin{equation}
\displaystyle u(x,t)=\mathrm{dn}(\xi)-2, \quad \xi=x-2t,
\end{equation}
and the graph of this function is
\newpage
\begin{figure}[!h]
\centering
\begin{center}
\includegraphics[angle=0,scale=.30]{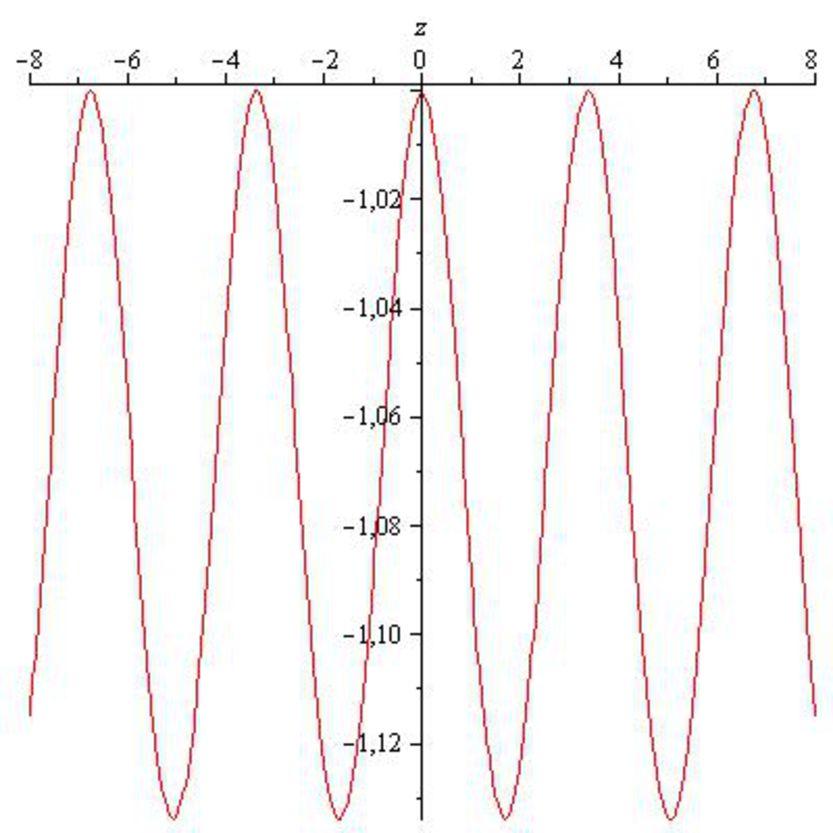}
\end{center}
\caption{Graph of Case $1.b$\, ($k=0.5$)}
\end{figure}

\noindent Note that by the choice of the parameters of this case the equation (\ref{F(f)}) becomes
$$
\displaystyle F(f)=-(f+3)(f+1)\Big(f-(-2+\frac{1}{2}\sqrt{3})\Big)\Big(f-(-2-\frac{1}{2}\sqrt{3})\Big).
$$
Since $F(f)$ has four different simple zeros, we expect periodic solution as in the graph.

\noindent \textbf{Case $2.a$ for $k=0$:}\quad The parameters are
$$
a=-2 , \quad b=\sqrt{3} , \quad c=1 , \quad \displaystyle d_1=\frac{3}{4} , \quad d_2=d_3=0 , \quad \beta=1.
$$
Hence the solution becomes
\begin{equation}
\displaystyle u(x,t)=\frac{1}{-2-\sqrt{3}\sin(\xi)},\quad \xi=x-t,
\end{equation}
and the graph of this function is
\begin{figure}[!h]
\centering
\begin{center}
\includegraphics[angle=0,scale=.30]{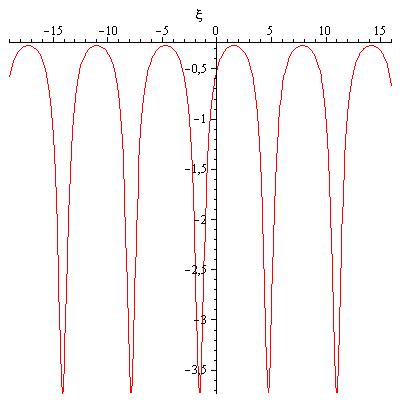}
\end{center}
\caption{Graph of Case $2.a$ \, ($k=0$)}
\end{figure}

\noindent Note that by the choice of the parameters of this case the equation (\ref{F(f)}) becomes
$$
\displaystyle F(f)=-f^2(f-(-2+\sqrt{3}))(f-(-2-\sqrt{3})).
$$
Here the function $F(f)$ has one double zero $f_2=0$ and two simple zeros $f_1=-2-\sqrt{3}$ and $f_3=-2+\sqrt{3}$
so $f_1<f_3<f_2$. As it is stated in section $2.2$, part $(4)$ we have periodic solution which can also be seen in the
above graph.

\noindent \textbf{Case $2.b$ for $k=0$:}\quad The parameters are chosen as
$$
a=2 , \quad \displaystyle b=-\sqrt{3}, \quad c=-1 , \quad d_1=\frac{3}{4}, \quad d_2=d_3=0 , \quad \beta=1.
$$
Hence the solution becomes
\begin{equation}
\displaystyle u(x,t)=\frac{1}{2-\sqrt{3}\cos(\xi)}, \quad \xi=x+t,
\end{equation}
and the graph of this function is
\begin{figure}[!h]
\centering
\begin{center}
\includegraphics[angle=0,scale=.30]{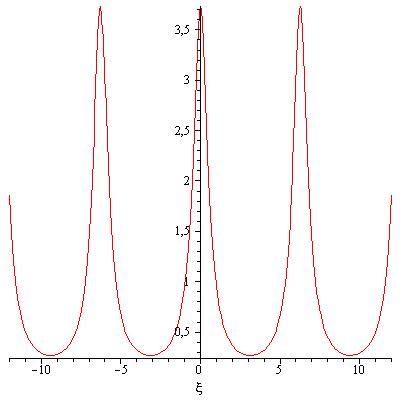}
\end{center}
\caption{Graph of Case $2.b$\, ($k=0$)}
\end{figure}

\noindent Note that by the choice of the parameters of this case the equation (\ref{F(f)}) becomes
$$
\displaystyle F(f)=-f^2(f-(2+\sqrt{3}))(f-(2-\sqrt{3})).
$$
The function $F(f)$ has one double zero $\displaystyle f_2=0$ and two simple zeros $f_1=2-\sqrt{3}$ and $f_3=2+\sqrt{3}$
so $f_2<f_1<f_3$. As it is given in section $2.2$, part $(4)$, the solution is periodic, which can be easily seen in the graph.

\noindent \textbf{Case $2.b$ and Case $2.c$ for $k=1$:}\quad The parameters are chosen as
$$
a=1 , \quad \displaystyle b=-\sqrt{\frac{7}{8}} , \quad c=-\frac{9}{2}, \quad d_1=10 , \quad d_2=4 , \quad d_3=-1 , \quad \beta=\sqrt{7}.
$$
Hence the solution becomes
\begin{equation}
\displaystyle u(x,t)=\frac{1}{1-\sqrt{\frac{7}{8}}\mathrm{sech}(\sqrt{7}\xi)}, \quad \xi=x+\frac{9}{2}t,
\end{equation}
\newpage
and the graph of this function is
\begin{figure}[!h]
\centering
\begin{center}
\includegraphics[angle=0,scale=.30]{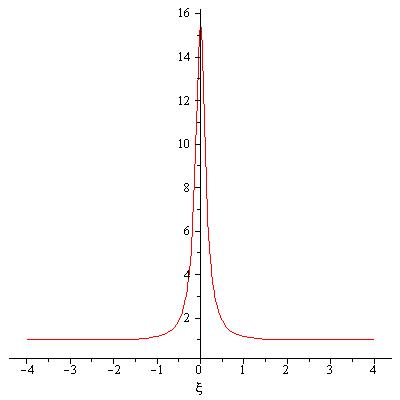}
\end{center}
\caption{Graph of Case $2.b$-$2.c$ \,($k=1$)}
\end{figure}

\noindent Note that by the choice of the parameters of this case the equation (\ref{F(f)}) becomes
$$
F(f)=-(f-(8+2\sqrt{14}))(f-(8-2\sqrt{14}))(f-1)^2.
$$
The numerical values of the zeros of $F(f)$ are such that the graph corresponds to the exact solitary wave solution given in
section $3.3$, part $(a)$.

\noindent \textbf{Case $2.e$ for $k=0$:}\quad The parameters are chosen as
$$
a=1 , \quad \displaystyle b=2 , \quad c=-\frac{1}{3} , \quad d_1=\frac{5}{18} , \quad d_2=-\frac{1}{6} , \quad d_3=\frac{1}{24} , \quad \beta=\frac{2\sqrt{3}}{3}.
$$
Hence the solution becomes
\begin{equation}
\displaystyle u(x,t)=\frac{\sin(\frac{2\sqrt{3}}{3}\xi)}{\sin(\frac{2\sqrt{3}}{3}\xi)+2}, \quad \xi=x+\frac{1}{3}t,
\end{equation}
and the graph of this function is
\begin{figure}[!h]
\centering
\begin{center}
\includegraphics[angle=0,scale=.30]{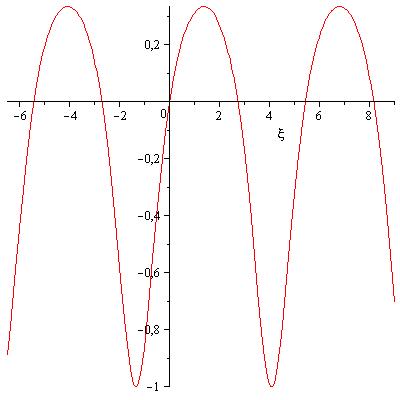}
\end{center}
\caption{Graph of Case $2.e$\, ($k=0$)}
\end{figure}

\noindent Note that by the choice of the parameters of this case the equation (\ref{F(f)}) becomes
$$
\displaystyle F(f)=-(f-1)^2(f+1)\Big(f-\frac{1}{3}\Big).
$$
Here the function $F(f)$ has one double zero $f_2=1$ and two simple zeros $f_1=-1$ and $f_3=\frac{1}{3}$
so $f_1<f_3<f_2$. As it is noted in section $2.2$, part $(4)$ we have periodic solution which can be seen in the graph.

\noindent \textbf{Case $2.f$ for $k=0$:}\quad The parameters are chosen as
$$
a=1 , \quad \displaystyle b=2 , \quad c=-\frac{1}{3}, \quad d_1=\frac{5}{18} , \quad d_2=-\frac{1}{6} , \quad d_3=\frac{1}{24} , \quad \beta=\frac{2\sqrt{3}}{3}.
$$
Hence the solution becomes
\begin{equation}
\displaystyle u(x,t)=\frac{\cos(\frac{2\sqrt{3}}{3}\xi)}{\cos(\frac{2\sqrt{3}}{3}\xi)+2}, \quad \xi=x+\frac{1}{3}t,
\end{equation}
and the graph of this function is
\begin{figure}[!h]
\centering
\begin{center}
\includegraphics[angle=0,scale=.30]{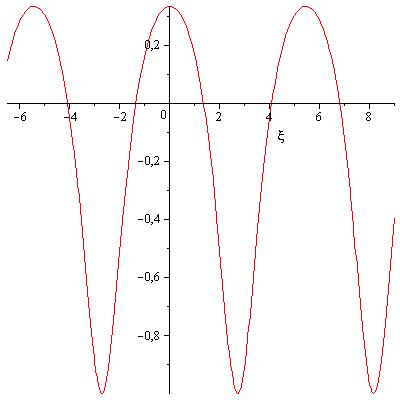}
\end{center}
\caption{Graph of Case $2.f$\, ($k=0$)}
\end{figure}

\noindent Note that by the choice of the parameters of this case the equation (\ref{F(f)}) becomes
$$
\displaystyle F(f)=-(f-1)^2(f+1)\Big(f-\frac{1}{3}\Big).
$$
The zeros of the function $F(f)$ are same as in the previous case. So the graph fits to the fact given in section $2.2$, part $(4)$.

\noindent \textbf{Case $2.f$ for $k=1$:}\quad The parameters are chosen as
$$
a=1,\quad \displaystyle b=2,\quad c=\frac{1}{6},\quad d_1=\frac{1}{9},\quad d_2=d_3=0,\quad \beta=\frac{\sqrt{3}}{3}.
$$
Hence the solution becomes
\begin{equation}
\displaystyle u(x,t)=\frac{\mathrm{sech}(\frac{\sqrt{3}}{3}\xi)}{\mathrm{sech}(\frac{\sqrt{3}}{3}\xi)+2}, \quad \xi=x-\frac{1}{6}t,
\end{equation}
\newpage

\noindent and the graph of this function is
\begin{figure}[!h]
\centering
\begin{center}
\includegraphics[angle=0,scale=.30]{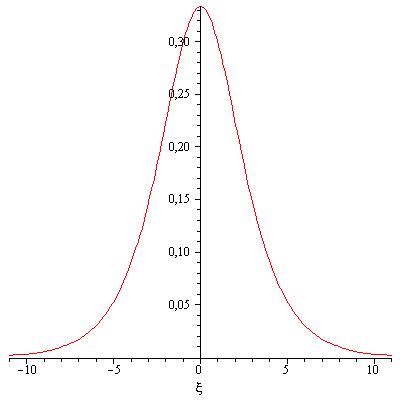}
\end{center}
\caption{Graph of Case $2.f$\, ($k=1$) }
\end{figure}

\noindent Note that by the choice of the parameters of this case the equation (\ref{F(f)}) becomes
$$
\displaystyle F(f)=-f^4-\frac{2}{3}f^3+\frac{1}{3}f^2=-\Big(f-\frac{1}{3}\Big)(f+1)f^2.
$$
The numerical values of the zeros of $F(f)$ are such that the graph corresponds to the exact solitary wave solution given in
section $3.3$, part $(a)$.

\section{Conclusion}
We have studied symmetry reduced (traveling waves) equations of the  Kaup-Boussinesq (KB) type of coupled degenerate KdV equations for $\ell=2$.
The reduced equation turns out to be such that the square of the derivative of the dependent variable is equal to a fourth degree polynomial of the dependent variable. There are four arbitrary constants in the polynomial function. We have investigated all possible cases and gave all solitary wave solutions which rapidly decay to some constants of the $(\ell=2)$ KB equations. There are periodic solutions of this set of coupled KdV equations in terms of the Jacobi elliptic functions. We first introduced special solutions of this type where the zeros of $F(f)$ satisfy certain constraints. If we remove these constraints among the zeros we obtained the most general solution in terms of the elliptic functions of KB system under the assumed symmetry. There are four different such solutions which differ by the initial values at the origin. For illustration we have given the graphs of some interesting solutions. We have also initiated the work on the cases for $\ell=3$ and $\ell=4$. We have given some results concerning these cases. A detailed study of the traveling wave solutions of the cases  $\ell=3$ and $\ell=4$ will be communicated later.

\section{Acknowledgment}
  This work is partially supported by the Scientific
and Technological Research Council of Turkey (T\"{U}B\.{I}TAK).

\end{document}